\documentclass[a4paper,12pt]{article}      
\usepackage{a4}
\usepackage{amsfonts}
\usepackage{amssymb}
\usepackage{cite}
\usepackage{epsfig}
\usepackage{psfig}
\usepackage[all]{xy}
\usepackage{amsmath}
\usepackage{cite}
\usepackage[dvips]{geometry}
\usepackage{lscape}

\author{}

\newcommand{\be}{\begin{equation}}
\newcommand{\ee}{\end{equation}}

\newcommand{\beq}{\begin{equation}}
\newcommand{\eeq}{\end{equation}}

\newcommand{\ba}{\begin{array}}
\newcommand{\ea}{\end{array}}
\newcommand{\bea}{\begin{eqnarray}}
\newcommand{\eea}{\end{eqnarray}}

\newcommand{\ov}{\overline}
\def\IR{\relax{\rm I\kern-.18em R}}

\def\IP{\relax{\rm I\kern-.18em P}}
\def\inbar{\vrule height1.5ex width.4pt depth0pt}
\def\IC{\relax\,\hbox{$\inbar\kern-.3em{\rm C}$}}

\def\K3{{\bf K3}}

\def\i{\iota}

\def\ov{\overline}

\def\n2d{\cN_{V^*}^{\otimes 2}}

\def\IC{\mathbb{C}}

\def\IR{\mathbb{R}}

\def\IP{\mathbb{P}}

\def\cN{{\mathcal N}}

\begin{document}

\title{
\begin{flushright} \vspace{-2cm}
{
\small{HD-THEP-10-18}
}
\end{flushright}
\vspace{4.0cm}
Lectures on F-theory compactifications and model building \\
}
\vspace{2.5cm}
\author{\small Timo Weigand}

\date{}

\maketitle

\begin{center}
\emph{ Institut f{\"u}r Theoretische Physik, Universit\"at Heidelberg,\\
   Philosophenweg 19, 69120 Heidelberg, Germany } \\
\vspace{0.2cm}

\tt{ t.weigand@thphys.uni-heidelberg.de}
\vspace{1.0cm}
\end{center}
\vspace{1.0cm}

\begin{abstract}
\noindent  

These lecture notes are devoted to formal and phenomenological aspects of F-theory. We begin with a pedagogical introduction to the general concepts of F-theory, covering 
 classic topics such as the connection to Type IIB orientifolds, the geometry of elliptic fibrations and the emergence of gauge groups, matter and Yukawa couplings.
 As a suitable framework for the construction of compact F-theory vacua we describe a special class of Weierstrass models called Tate models, whose local properties are captured by the spectral cover construction. Armed with this technology we proceed with a survey of F-theory GUT models, aiming at an overview of basic conceptual and phenomenological aspects, in particular in connection with GUT breaking via hypercharge flux.

\end{abstract}

\thispagestyle{empty}
\clearpage

\tableofcontents

\section{Introduction}

String phenomenology is the branch of string theory that takes the theory seriously as a consistent, fundamental description of gravitational and particle interactions in four dimensions.
Its primary goal is to understand the solutions of this theory and their implications for physics in four dimensions.
This includes the investigation of both the mathematical structure of the space of string vacua and of their concrete particle phenomenological and cosmological properties.

In the perturbative region of the string landscape, two corners exhibit particularly attractive four-dimensional solutions:
The heterotic string heavily exploits the availability of exceptional gauge symmetry, into which the observed gauge group of the Standard Model can be embedded in elegant ways realising the idea of a grand unified theory (GUT).  Many different techniques have been developed to study these solutions in various regions of the moduli space, ranging from heterotic orbifolds and free-fermionic constructions to smooth Calabi-Yau compactifications with vector bundles, see e.g.\ the review \cite{Nilles:2008gq} and references therein.
Since both gravity and gauge dynamics descend from the closed string sector, all aspects of four-dimensional physics are sensitive to the global structure of the compactification space.

Perturbative Type II orientifolds with D-branes, on the other hand, are based on the classical gauge groups $U(N)$, $SO(N)$ and $Sp(2N)$. This type of constructions can directly accomodate the Standard Model gauge group, while GUT physics is not immediate.
The huge literature on Type IIA orientifolds with intersecting D6-branes and on Type IIB orientifolds with D7/D3-branes (reviewed e.g.\ in \cite{Lust:2004ks,Blumenhagen:2005mu,Blumenhagen:2006ci}) exploits furthermore the localisation of the gauge degrees of freedom along the D-branes, which are therefore of a different physical origin than the gravity sector in the bulk. This opens up the possibility of exploring a certain subclass of phenomenological questions in the context of local models, while other aspects of the associated four-dimensional physics remain sensitive to the full compactification details.

The strict separation between the phenomenological properties of the heterotic and brane constructions ceases to exist at finite values of the string coupling $g_s$. In fact, in all but a minute class of brane constructions $g_s$ is dynamical and varies over the compactification space in such a way as to leave the perturbative regime somewhere. The correct way to describe Type IIB compactifications with 7-branes in this generic situation is called F-theory \cite{Vafa:1996xn}. 
F-theory is a fascinating subject by itself because it geometrises the backreaction of the 7-branes on the ambient space and is therefore, in some sense, \emph{the} way to think about 7-branes. It incorporates certain strong coupling effects with breath-taking elegance. The rich and by now classic literature on the more formal aspects of F-theory reflects the amount to which these non-perturbative phenomena have mesmerized string theorists.

More recently it has been exploited that the four-dimensional solutions of F-theory are also interesting from a phenomenological viewpoint  \cite{Donagi:2008ca,Beasley:2008dc,Beasley:2008kw,Donagi:2008kj}. 
This is because at strong coupling new degrees of freedom - (p,q) strings - become light and realise exceptional gauge symmetries even in a theory based on branes.
The perturbative dichotomy between localisation of gauge degrees of freedom and exceptional gauge dynamics is therefore resolved. 
This bears exciting prospects for GUT model building and has been the subject of fruitful and intensive investigations in recent times. 
Most importantly, F-theory compactifications also inherit the favourable properties of M-theory and of Type IIB orientifolds with respect to moduli stabilisation: The combination of 3-brane instanton effects and background fluxes allows in principle for the stabilisation of K\"ahler and complex structure moduli within the framework of warped Calabi-Yau compactifications. 
This is motivation enough to take F-theory seriously as a promising corner for string phenomenology.

When dealing with F-theory compactifications one must be aware that to date there exists no description of F-theory  as a fundamental theory. In this respect F-theory has a status very different from M-theory, which  can - at least conjecturally  - be conceived as the theory reducing in its long-wavelength limit  to eleven-dimensional supergravity coupled to membranes. 
Early ideas \cite{Vafa:1996xn,Tseytlin:1996it,Hull:1995xh} to define an analogous twelve-dimensional theory whose fundamental objects are 3-branes have not lead to a consistent picture.
Rather F-theory should be thought of as a genuinely non-perturbative description of a class of string vacua which, in certain limits, is accessible by string dualities from three different corners of the M-theory star.
These are
\begin{itemize}
\item F-theory as (strongly coupled) Type IIB theory with 7-branes  and varying dilaton,
\item F-theory as dual to $E_8 \times E_8$ heterotic theory,
\item F-theory as dual to M-theory on a vanishing $T^2$.
\end{itemize}
Of these three, the F/M-theory duality probably captures the dynamics in the most general way. 
While particularly fruitful for concrete applications, the first two dualities can sometimes be misleading because they describe only certain aspects of the dynamics of a typical F-theory compactification.

These lectures intend to provide a pedagogical introduction to some of the technology and the phenomenological applications of F-theory model building.
A complimentary set of lecture notes covering general aspects of F-theory is \cite{Denef:2008wq}, while the specifics of F-theory GUT model building are surveyed in the review \cite{Heckman:2010bq}.

Section 2 aims at a pedagogical introduction to the main concepts of F-theory.
In section \ref{need} we begin with a definition of F-theory from the perspective of Type IIB orientifolds with 7-branes. The F/M-theory duality is briefly sketched in section \ref{sec_FM}. Given the geometric nature of F-theory, the bread and butter of the business is to understand the geometry of elliptic fibrations, which we approach in an elementary manner in section \ref{sec_Gen}. The connection to the Type II orientifold picture is made via Sen's orientifold limit in section \ref{sec_Sen}. A closer investigation of the appearance of gauge degrees of freedom from the singularities of the elliptic fibration concludes our first encounter with F-theory in section \ref{sec_Gau}. 

In section \ref{Technology} we introduce some of the more advanced technology of F-theory compactifications.  The basic algorithm to read off the gauge groups from a given model is presented in section \ref{sec_Tat}. The corresponding Tate models represent a convenient framework for F-theory compactifications.  Besides the pure geometry, gauge flux is an essential ingredient in  F-theory models (see section \ref{sec_Flu}). Note that it is this aspect that is currently the least understood. An account of charged matter and Yukawa couplings follows in section \ref{matter-gen}. 
In section \ref{Fhet} we recall basic aspects of F-theory/heterotic duality; an important ingredient  is the construction of vector bundles on elliptically fibered Calabi-Yau 3-folds via the spectral cover construction. 
Section \ref{spectral_cover} concludes our tour through the model building rules of F-theory with a review of the recent application of the spectral cover construction in general F-theory models with no heterotic dual.

In the final part of these lecture notes we apply what we have learned in sections \ref{sec_GenF} and \ref{Technology}  to the construction of $SU(5)$ GUT vacua. This is by now a vast and dynamical field, and rather than aiming at completeness we focus on some fundamental aspects. After briefly describing the decoupling idea underlying local models in section \ref{sec_SU(5)}, we investigate various options for GUT symmetry breaking in section \ref{GUTbreaking}. GUT breaking by hypercharge flux is critically assessed in section \ref{sec_Som}. The \emph{experimentum crucis} in GUT model building is proton decay, and we explain some of the challenges within the F-theory context in section \ref{proton}. An incomplete list of further phenomenological topics can be found in \ref{sec_Fur}.

\section{A first encounter with F-theory}
\label{sec_GenF}

\subsection{The need for a non-perturbative formulation of Type IIB with 7-branes}
\label{need}

\subsubsection*{Backreaction from 7-branes}

In this section we approach F-theory as the  strong coupling limit of Type IIB orientifolds with O7/O3-planes and D7/D3-branes. For background on Type II orientifolds we refer to existing reviews such as \cite{Lust:2004ks,Blumenhagen:2005mu,Blumenhagen:2006ci}.
The usual philosophy in the description of perturbative Type II orientifolds with D-branes is to neglect the backreaction of the branes and the orientifold planes on the background geometry in the spirit of a probe approximation. This is justified as long as asymptotically away from the brane the backreaction becomes negligible.
In this case, one can consider a large volume limit in which knowledge of the detailed form of the solution is not required at least to understand the main properties of the string vacuum.

To see when this approximation is justified, we consider a $p$-brane in ten dimensions. It represents a source term in the normal $n=9-p$ spatial directions. At a heuristic level, this leads to a Poisson equation for the background fields sourced by the brane.  Schematically denoting these sourced fields as $\Phi$, one can write this as
\bea
\Delta \Phi (r) \simeq \delta (r) \, \Longrightarrow \, \Phi(r) \simeq \frac{1}{r^{n-2}} \quad\quad {n >2}.
\eea
More precisely, within Type II supergravity  the BPS solution for a stack of $N$ $p$-branes along directions $\mu=0,1 \ldots, p$, $p < 7$, takes the form \cite{Horowitz}
\bea
&& ds^2 = H_p^{-1/2} \eta_{\mu \nu} dx^{\mu} dx^{\nu} + H_p^{1/2} \sum_i dx^i dx^i, \\
&& e^{2 \phi} = e^{2 \phi_0}  H_p^{\frac{3-p}{2}}, \quad C_{p+1} = \frac{H_p^{-1} -1}{e^{\phi_0}} dx^0 \wedge \ldots \wedge dx^p, \\
&& H_p = 1 +\Big( \frac{r_p}{r} \Big)^{(7-p)} , \quad\quad r_p^{(7-p)} = {\rm const} \, e^{\phi_0} \, N. 
\eea
Here $\phi$ denotes the ten-dimensional dilaton and the Ramond-Ramond $(p+1)$-form potential $C_{p+1}$ couples electrically to the $p$-brane.
The backreaction is governed by the harmonic function $H_p$, which asymptotes to unity away from the brane. In particular, $e^{\phi_0}$ denotes the asymptotic value of $e^{\phi}$ and can be taken as the value of the string coupling $g_s$ relevant in the large volume limit.

However, the above logic goes through only if the codimension of the brane $n > 2$. The critical case $n=2$ corresponds precisely to D7-branes in Type IIB theory.
A D7-brane along, say, dimensions $0,\ldots, 7$  looks like a charged point particle localised in the two normal directions $8,9$ - a cosmic string \cite{Greene:1989ya}. Solutions to the two-dimensional Poisson equation  scale \emph{logarithmically} with the distance to the source. Such a logarithmic profile is in sharp contrast with the favourable asymptotics for lower dimensional branes encountered above.

Let us see how this heuristic argument applies in more detail. Recall, e.g.\ from \cite{Blumenhagen:2006ci},  that the string frame Type IIB effective action in the democratic formulation is given by
\bea
\label{String-action}
S^{(S)}_{IIB}& = & \frac{2 \pi}{\ell_s^8} \Big( \int d^{10}x\ e^{-2 \phi} ( \sqrt{-g} R + 4 \partial_M \phi \, \partial^M \phi) -\frac{1}{2} e^{-2 \phi}  \int H_3 \wedge \ast H_3  \nonumber \\ 
&& -  \frac{1}{4} \sum_{p=0}^4  \int F_{2p+1}  \wedge \ast F_{2p+1} - \frac{1}{2} \int C_4 \wedge H_3 \wedge F_3      \Big).
\eea
Here $\ell_s = 2 \pi \sqrt{\alpha'}$ and the field strengths are defined as
\bea
&& H_3 = dB_2, \quad F_1 = dC_0, \quad F_3 = dC_2 - C_0 \, dB_2, \nonumber \\ 
&& F_5 = dC_4 - \frac12 C_2 \wedge dB_2 + \frac12 B_2 \wedge dC_2,
\eea
supplemented by the duality relations $F_9 = \ast F_1$, $F_7 = - \ast F_3$, $F_5 = \ast F_5$  at the level of equations of motion.

The RR-field sourced electrically by a D7-brane is $C_8$ and is dual to the axion $C_0$ that combines with the string coupling $g_s=e^{\phi}$ into the complex axio-dilaton 
\bea
\tau = C_0 +\frac{i}{g_s} .
\eea
The D7-brane action is the sum of the two terms
\bea
\label{S-CS}
& S_{\rm DBI} &= - \frac{2\pi}{\ell_s^8} \int d^7 \zeta \, e^{- \phi} \, \sqrt{ {\rm det} (g + {2 \pi \alpha' {\cal F}}/{2 \pi})} , \\
&S_{\rm CS}& = - \frac{2\pi}{\ell_s^8}  \int  \, {\rm tr} \, \, {\rm exp}(2 \pi \alpha' {\cal F})\, \sum_p C_{2p} \,   \,  \sqrt{\frac{\hat A(T)}{\hat A(N)}},
\eea 
where  $2 \pi \alpha' {\cal F} = 2 \pi \alpha' {F} +B_2$ in terms of the Yang-Mills field strength $F$ and the last factor denotes curvature contribution in terms of the A-roof-genus (see e.g. \cite{Nakahara:2003nw}) of the tangent and normal bundle to the D7-brane.
Define the complex coordinate $z= x^8 + i x^9$ for the dimensions perpendicular to the 7-brane. Taking into account constraints from supersymmetry, it turns out that the axio-dilaton must be a holomorphic function in $z$.
Therefore the Poisson equation for $C_8$ in presence of a 7-brane at $z=z_0$ takes the form
\bea
\label{F9}
d \ast F_9 = \delta^{(2)}(z- z_0).
\eea 
The integrated form of (\ref{F9}) is 
\bea
1 =  \int_{\mathbb C} d \ast F_9 = \oint_{S^1} F_1 =   \oint_{S^1} d C_0. 
\eea
A simple solution can be found that is valid close to the brane at $z=z_0$, 
\bea
\label{NP-Equ-Gauss}
\tau(z)  = \tau_0 + \frac{1}{2 \pi i}{\rm ln} (z-z_0)  + \ldots, \quad\quad\quad
\eea 
where we have omitted possible regular terms in $z$. Note in particular that $g_s \rightarrow 0$ at the position of the brane $z=z_0$.\footnote{This does not mean that the gauge theory on the 7-brane is trival: The relevant frame is the Einstein frame, where the 7-brane gauge coupling is independent of $g_s$ and given to leading order by the volume of the internal 4-cycle wrapped by the brane (for the case of compactification to 4 dimensions).}

Far away from the brane, the simple solution of $\tau$ must be modified. 
Nonetheless we see already that the background exhibits a logarithmic branch cut in the complex plane normal to the D7-brane.
A careful analysis of the Einstein equations reveals  \cite{Greene:1989ya} that asymptotically away from the brane, spacetime becomes locally flat, but suffers a deficit angle. Contrary to the backreaction of lower-dimensional branes, this effect does not asymptote away and the probe philosophy is, strictly speaking, not applicable in this region.

Despite the non-trivial deficit angle at infinity, one can identify a region where the backreaction of the brane on the geometry is negligible. Following the   lucid discussion in \cite{Braun:2008ua}
one can, ignoring the higher terms, rewrite the solution (\ref{NP-Equ-Gauss}) as
\bea
\label{lambda}
\tau(z) = \frac{1}{2 \pi i} \, {\rm ln} \frac{z-z_0}{\lambda} \quad \Longrightarrow \quad e^{-\phi} =  - \frac{1}{2\pi} \, {\rm ln} |\frac{z-z_0}{\lambda}|
\eea
in the vicinity of the brane, where $\lambda$ is related to $\tau_0$. At the point $z- z_0 = \lambda$ one encounters $g_s  = e^{\phi} \rightarrow \infty$. The presence of this special point breaks the naively expected rotational invariance around the source, a clear sign of  backreaction. However, for $|z-z_0| \ll |\lambda|$ the geometry is approximately flat. It is this region that is the analogue of the asymptotic large distance limit where the backreaction of $p$-branes with $p < 7$ discussed above is negligible.
The limit of weak coupling is the one where this region (and distances therein) is large enough to trust effective supergravity, i.e.\ where $\lambda \rightarrow \infty$.\footnote{In compactifications,
while the profile of $\tau$ is set by the brane configurations, the integration constant $\tau_0$ or overall scale $\lambda$ remains a modulus of the low-energy effective theory.} I.e.\ in this limit the inevitable increase of $g_s$ away from the brane happens at larger and larger distances, and one remains at weak coupling as long as one focuses on suitable radius around the brane.

Generically, however, the fact that $g_s$ develops a strongly varying non-trivial profile obscures an interpretation of the background in terms of perturbation theory. As just discussed, in generic situations $\tau$ inevitably takes values of order 1 and larger in some regions of the compactification manifold.
Besides, the expression for $\tau$ must receive strong stringy corrections in the vicinity of an orientifold-plane, which carries $-4$ units of $C_8$-charge. A naive solution
$\tau(z)  \simeq  - \frac{4}{2 \pi i}{\rm ln} (z-z_0)$ which would be suggested by the above arguments is unphysical as the dilaton would become negative close to the O-plane.

An exception to the inevitable variation of $\tau$ in the presence of 7-branes and O-planes is the very special situation where all 7-brane charges cancel \emph{locally} because a suitable number of D7-branes and O7-planes lie on top of each other. 
In this case, $g_s$ is constant on the entire compactification manifold and in absence of stabilising effects such as fluxes it can freely be chosen to lie in the perturbative window $g_s \ll 1$ everywhere. In fact, one should reserve the term 'perturbative' for this non-generic configuration.
This establishes that a proper treatment of backgrounds with varying axio-dilaton is not an option, but rather required for a satisfactory definition of generic configurations with 7-branes.

\subsubsection*{Monodromies and $SL(2,\mathbb Z)$ invariance}

A peculiar feature of the 7-brane backreaction is the appearance of monodromies.
The monodromy associated with the simple solution (\ref{NP-Equ-Gauss}) implies that as one encircles the position of the D7-brane in the $z$-plane, the axio-dilaton transforms as
\bea
\label{tauplus1}
\tau \rightarrow \tau +1.
\eea
At first sight this might come as a shock as it seems to make a consistent interpretation of the background solution impossible.
The \emph{deus ex machina} approaching to our rescue is the fact that Type IIB is invariant under  $SL(2,\mathbb Z)$ transformations, of which (\ref{tauplus1}) is the simplest example.
As is most readily verified after transforming (\ref{String-action}) into Einstein frame (see e.g.\  \cite{Blumenhagen:2006ci}), 
the classical Type IIB action enjoys the $SL(2,\mathbb R)$ invariance
 \bea
 \label{SL2R}
   \small{   \tau \rightarrow \frac{a \tau + b}{ c \tau + d}, \quad\quad \left( \begin{array}{c}
            C_2 \\
            B_2  \end{array} \right)  }     \rightarrow  \small{ \left( \begin{array}{c}
a C_2 + b B_2 \\
c C_2  + d B_2 \end{array} \right)  = M  \left( \begin{array}{c}
            C_2 \\
            B_2  \end{array} \right)  }  , \quad  \quad {\rm det} M=1.\label{SL2Z}  \eea
Note that $C_4$ is invariant under these transformations.
This classical symmetry is broken at the non-perturbative level to $SL(2,\mathbb Z)$. The reason is that $D(-1)$ instanton effects involve a factor $e^{2 \pi i \tau}$. Invariance of such quantum effects under transformations of the type 
  \[   \small{  M = \left( \begin{array}{cc}
            1 & b  \\
             0 & 1 \end{array} \right):  \quad C_0 \rightarrow C_0 + b }    \]
 restricts $b \in \mathbb Z$, thus reducing $SL(2,\mathbb R)$ to $SL(2,\mathbb Z)$.

Therefore as one encircles a D7-brane, the full background transforms by the $SL(2,\mathbb Z)$ action $T=\tiny{\begin{pmatrix} 1 & 1  \\  0 & 1    \end{pmatrix}}$, and the monodromy action is merely a symmetry of the theory.

\subsubsection*{$[p,q]$-branes}

Once we have accepted that a consistent interpretation of D7-branes forces us to take the  $SL(2,\mathbb Z)$ symmetry at face value, we are lead to more exotic objects than D7-branes which have no interpretation in terms of perturbation theory. The new objects we must include are $[p,q]$-branes and corresponding $\tiny{ \begin{pmatrix} p   \\  q    \end{pmatrix} }$ strings \cite{Schwarz:1995dk}.
To see this, recall from the Polyakov worldsheet action that the fundamental superstring is charged electrically under the NS-NS $B_2$-field. Given the mixing of $B_2$  with the RR 2-form $C_2$ under general $SL(2,\mathbb Z)$ transformations as in (\ref{SL2R}) there must also exist an analogous object charged electrically under $C_2$. This, of course, is nothing other than the D1-string, and its corresponding coupling to $C_2$ is the Chern-Simons action. It is therefore appropriate to combine the F1- and the D1-string into an $SL(2,\mathbb Z)$ doublet: An F1-string is represented as the vector  $\tiny{\begin{pmatrix} 1   \\  0    \end{pmatrix}}$  and a D1-string is a ${\tiny \begin{pmatrix} 0   \\  1    \end{pmatrix}}$ object. A general $\tiny{ \begin{pmatrix} p   \\  q    \end{pmatrix} }$ string carries $p$ units of electric $B_2$-charge and $q$ units of electric $C_2$-charge. Such objects exist as supersymmetric bound states for $p, q$ co-prime \cite{Witten:1995im}. 
Note that in perturbative Type IIB theory only  ${\tiny\begin{pmatrix} 1   \\  0    \end{pmatrix}}$ strings are present as the fundamental objects, while the D1-string enters as a solitonic, non-perturbative object. 

In perturbative Type IIB theory, a D7-brane is by definition a hypersurface on which fundamental strings can end. This motivates the definition, in strongly-coupled Type IIB theory, of a $[p,q]$-7-brane as the hypersurface on which ${\tiny \begin{pmatrix} p   \\  q    \end{pmatrix}}$ strings can end. 
Two $[p,q]$-7-branes of different $p,q$ are called mutually non-local.

A general $[p,q]$-brane induces  an $SL(2,\mathbb Z)$ monodromy $M_{p,q}$ on the background fields as one encircles the location of the brane. This monodromy generalises the perturbative action (\ref{tauplus1}) induced by $T=M_{1,0}$ and can be shown to take the form
\bea
\label{Mpq}
M_{p,q} = g_{p,q} \, \, M_{1,0}  \, \, g_{p,q}^{-1} = {\begin{pmatrix} 1-pq & p^2  \\  -q^2 & 1+pq    \end{pmatrix}}.
\eea
Here  $g_{p,q} =\tiny{\begin{pmatrix} p & r  \\  q & s    \end{pmatrix}} $ is the $SL(2,\mathbb Z)$ matrix that transforms a  $\tiny{ \begin{pmatrix} 1   \\  0    \end{pmatrix} }$ string into a  $\tiny{ \begin{pmatrix} p   \\  q    \end{pmatrix} }$ string.\footnote{Note that $r$ and $s$ are not uniquely determined. This ambiguity drops out in all physically relevant quantities.} 
The only eigenvector of $M_{p,q}$ is a $\tiny{\begin{pmatrix} p   \\  q   \end{pmatrix}}$-string itself, any other type of string gets transformed as transported around the location of the $[p,q]$-7-brane.

Every $[p,q]$-brane as such can be mapped into a $[1,0]$-brane, i.e.\ a conventional D7-brane, by an $SL(2,\mathbb Z)$ transformation. Locally around each single 7-brane the geometry therefore is indistinguishable from the one  backreacted  by a D7-brane. However, in the presence of mutually non-local $[p,q]$-branes in the above sense, new phenomena arise because the various branes cannot be \emph{simultaneously} transformed into a D7-brane. And, most importantly, a consistent compactification necessarily includes 7-branes of different type. 

The simplest example where this becomes apparent starts from a perturbative Type IIB orientifold on $T^2$ modded out by $\Omega (-1)^{F_L} \sigma$. The geometric orientifold action $\sigma$ transforms the complex coordinate of the torus as $z\rightarrow -z$. Its four fixed points are the location of O7-branes. Local tadpole cancellation in perturbative models (in the sense introduced above) requires 4 D7-branes (plus their image branes) on top of each O7-plane, resulting in a famous $SO(8)^4$ gauge group.
The $[p,q]$-brane interpretation of this configuration was given by Sen \cite{Sen:1996vd}, who showed that eight-dimensional Type II compactifications require a set of three different $[p,q]$-branes with $[p,q]$-labels
\bea
A: [1,0], \quad\quad B: [3,-1], \quad\quad C: [1,-1].
\eea
In this notation $A$-type branes correspond to perturbative D7-branes.  In fact the branes $B$ and $C$ are chosen such that the combined monodromy $M_{BC} = M_{3,-1} \, M_{1,-1}$ acts on a  $\tiny{\begin{pmatrix} 1   \\  0   \end{pmatrix}}$ string by orientation reversal,
\bea
\label{Oaction}
M_{3,-1} \, M_{1,-1} {\begin{pmatrix} 1   \\  0   \end{pmatrix}} = -  {\begin{pmatrix} 1   \\  0   \end{pmatrix}},
\eea
as in the context of an orientifold theory.
This identifies the $BC$-system as a perturbative O7-plane.

In generic configurations, however, objects with general $SL(2,\mathbb Z)$ monodromies will play a role, simply because the action of $M_{1,0}$ and $M_{BC}$ does not generate the full $SL(2,\mathbb Z)$.
This raises a puzzle: What are the consistency conditions for theories with mutually non-local $[p,q]$-branes? Is any configuration of coincident branes like the $BC$ system allowed?
Given the dyonic nature of $[p,q]$-branes, perturbative methods are bound to fail. For compactifications to eight dimensions one might try to keep track of the $SL(2,\mathbb Z)$ monodromies by hand and search for intrinsically consistent configurations, but already in six dimensions this becomes intractable.

It is therefore quite remarkable that there exists a reformulation of the problem that allows us to almost blindly read off the consistent configurations of 7-branes. This formulation is the much sought-after F-theory.

\subsubsection*{Towards a geometric description}

The crucial insight \cite{Vafa:1996xn} that underlies the formulation of such a non-perturbative theory is the identification of the $SL(2,\mathbb Z)$ symmetry of ten-dimensional Type IIB supergravity with the geometric $SL(2,\mathbb Z)$ action on the complex structure of a two-torus $T^2$. 
Inspired by the $SL(2,\mathbb Z)$ transformation of $\tau$ one interprets the axio-dilaton as the complex structure of a ficticious elliptic curve. 
The variation of $\tau$ in presence of a set of 7-branes is therefore modelled as the variation of the complex structure of an elliptic curve transverse to  the location of the 7-branes. Such a structure defines an elliptic fibration, and the non-triviality of the fibration is a measure for how strongly the axio-dilaton varies as a consequence of the backreaction of the branes.

The F-theory conjecture states that the physics of Type IIB orientifold compacitifications with 7-branes on the complex n-fold $B_n$ is encoded in the geometry of an (n+1)-fold $Y_{n+1}$ which is elliptically fibered over $B_n$. The elliptic fiber itself is not part of the physical spacetime but merely a  book-keeping device that accounts for the variation of $\tau$. In particular, at the location of 7-branes the axio-dilaton $\tau$ diverges as seen for the solution (\ref{NP-Equ-Gauss}). If the complex structure of an elliptic curves diverges, this indicates the shrinking of a one-cycle and thus the degeneration of the elliptic curve. Thus the degeneration locus of the elliptic curve describes the presence of 7-branes.

In fact, duality with M-theory yields additional restrictions on the relevant elliptic fibrations. First, for ${\cal N}=1$ supersymmetry to be conserved, the space $Y_{n+1}$ has to be Calabi-Yau. Further, only the limit of vanishing elliptic fiber is to be considered. Both these facts will become apparent in the context of F/M-theory  duality reviewed in the next section.

To return to the simple example alluded to before, understanding compactifications of Type IIB orientifolds to eight dimensions involves F-theory on a Calabi-Yau 2-fold $Y_2$. In this case there is only one such Ricci flat 2-fold, the famous $K3$. The physically most interesting situation of course corresponds to F-theory on an elliptically fibered 4-fold $Y_4$, which is supposed to capture Type IIB orientifolds compactified to four dimensions on the 3-fold $B_3$.

\subsection{F/M-theory duality and Calabi-Yau 4-folds}
\label{sec_FM}

So far we have motivated F-theory as a clever way to think about Type IIB  orientifolds with 7-branes.
To determine the details of the effective action and of the physical degrees of freedom though, it is useful to approach F-theory via duality with eleven-dimensional M-theory, taken as its long-wavelength limit of eleven-dimensional supergravity coupled to M2/M5-branes. This duality also provides a more direct way to uncover the appearance of the elliptic curve in F-theory.

The starting point is therefore compactification of eleven-dimensional supergravity on $\mathbb R^{1,9} \times T^2$. Let $\tau$ be the complex structure of the torus $T^2 = S^1_A \times S^1_B$. E.g.\ for the special case of a rectangular torus we have $\tau  = i \, \frac{R_{A}}{R_B}$ in terms of the radii $R_A$ and $R_B$ of $S^1_A$ and $S^1_B$. The M/F-theory duality consists in taking the limit of vanishing torus volume in the following two-step procedure \cite{Witten:1996bn}:
\begin{itemize}
\item
The circle $S^1_A$  with radius $R_{A}$ is identified as the M-theory circle in the reduction from M-theory to Type IIA theory in ten dimensions so that $g_{IIA} \simeq \frac{R_{A}}{\ell_s}$. The limit $R_{A} \rightarrow 0$ therefore corresponds to taking the perturbative IIA limit of M-theory, and we recover weakly coupled Type IIA theory on $\mathbb R^{1,9} \times S_B^1$.
\item
Now perform a T-duality along the remaining circle $S_B^1$ with radius $R_{B}$. This yields Type IIB theory on $\mathbb R^{1,9} \times \tilde S_B^1$ with dual radius  $\tilde R_{B} = \frac{\ell_s^2}{R_B}$. The limit $R_B \rightarrow 0$ results in ten-dimensional Type IIB theory.  
\end{itemize}
Application of the  T-duality transformation identifies the IIB coupling  as $g_{IIB} \simeq g_{IIA}  \frac{\ell_s}{R_B} \simeq  \frac{R_{A}}{R_B} \simeq {\rm Im}(\tau)$. The last step applies to rectangular tori but has simple generalisations. A more precise analysis as worked out in detail e.g.\ in \cite{Denef:2008wq} allows one to trace back the Type IIB  RR axion $C_0$ to the type IIA RR one-form $C_1$. The latter in turn derives from the real part of the complex structure of the M-theory torus.

What we have sketched is a description of Type IIB theory in ten dimensions in terms of M-theory on $\mathbb R^{1,9} \times T^2$, where the IIB axio-dilaton  $\tau = C_0 + \frac{i}{g_s}$ is identified with the complex structure of the M-theory $T^2$. F-theory on $T^2$ can therefore be defined as the ten-dimensional IIB theory which is dual, in the above sense, to M-theory on $\mathbb R^{1,9} \times T^2$.
 Note that for the duality to work the volume of the M-theory $T^2$ has to vanish. This is the physical reason why only the complex structure $\tau$, but not the volume of the elliptic curve appears as a physical field in Type IIB/F-theory.

The above logic 
  extends adiabatically to the more general case of a non-trivial elliptic fibration rather than a direct product $\mathbb R^{1,9} \times T^2$.
The physically most interesting situation is of course that of M-theory compactification on $\mathbb R^{1,2} \times Y_{4}$, where $Y_{4}: T^2 \rightarrow B_3$ is  a complex 4-fold elliptically fibered over a 3-complex dimensional base $B_3$. If the 4-fold $Y_4$ is Calabi-Yau this yields a three-dimensional effective theory  with four supercharges\cite{Becker:1996gj,Dasgupta:1999ss,Haack:2001jz}.
In the limit of vanishing fiber volume the fourth dimension grows large and this setup is dual to the four-dimensional effective theory obtained by Type IIB compactification on $B_3$, or, by definition, F-theory on $Y_4$. The four supercharges lead to an  ${\cal N}=1$ effective theory in four dimensions.
Thus the Calabi-Yau property of $Y_4$ is an obvious requirement from the M-theory perspective.

Kaluza-Klein reduction of the M-theory 3-form allows one to recover also the higher Type IIB RR-forms.
We denote by $\alpha$ and $\beta$ the periodic coordinates along the M-theory torus $T^2  = S_A^1 \times S_B^2$  and focus on a generic non-singular fiber for the time being.
Reduction of $C_3$ yields
\bea
\label{C3}
C_3 = \tilde C_3 + B_2 \wedge d\alpha + C_2 \wedge d\beta + B_1 \wedge d \alpha \wedge d\beta.
\eea
After T-duality along $\beta$ and decompactification to four dimensions, $\tilde C_3$  furnishes the degrees of freedom of the RR 4-form $C_4 = \tilde C_3 \wedge d \beta$, while $B_1$ becomes part of the four-dimensional metric $g_{i \beta}$. $B_2$ and $C_2$, on the other hand, translate into the NS-NS and R-R 2-forms in Type IIB. Their transformation properties as a doublet under the $SL(2, \mathbb Z)$ group  derive immediately from the geometric $SL(2, \mathbb Z)$ transformation of the $A$ and $B$ cycle of the M-theory torus.

It is beyond the scope of these lectures to enter a detailed discussion of the F-theory effective action or of the various subtleties of the F-theory limit starting from the M-theory reduction on an elliptic fibration. An in-depth derivation of the four-dimensional F-theory effective action via M/F-theory duality has been provided in \cite{Eff-action} (see also \cite{Denef:2008wq}), which we recommend for more details.
An important ingredient in Kaluza-Klein reduction, e.g.\ to determine the precise massless supergravity spectrum, are the topological properties of Calabi-Yau 4-folds. Many aspects of the geometry and topology of (elliptic) Calabi-Yau 4-folds can be found e.g. in \cite{Mayr:1996sh,Klemm:1996ts}. Suffice it here to recall for completeness that the Hodge diagram of Calabi-Yau 4-folds is characterised by three independent Hodge numbers $h^{1,1}$, $h^{2,1}$, $h^{3,1}$, to which $h^{2,2}$ is related via
\bea
h^{2,2} = 2\,  (22 + 2 h^{1,1} + 2 h^{3,1} - h^{2,1}).
\eea
An important quantity for model building, which will appear prominently in the sequel, is the Euler characteristic 
\bea
\label{eq-Euler1}
\chi(Y_4) = \int_{Y_4} c_4(T_{Y_4}) =  6 \,  (8 +  h^{1,1} +  h^{3,1} - h^{2,1}).
\eea

\subsection{The geometry of elliptic fibrations}
\label{sec_Gen}

The above considerations show that understanding the non-perturbative F-theoretic region of the string landscape requires familiarity with the concept of elliptic fibrations, to which we now turn in some detail. The problem splits into two parts: We first need to understand elliptic curves as such, and then find a way to describe their fibration over the base  $B_n$ such as to form an elliptically fibered Calabi-Yau manifold $Y_{n+1}$,
\bea
\pi: Y_{n+1} \rightarrow B_n .
\eea

\subsubsection*{Elliptic curves as ${\mathbb P}_{2,3,1}[6]$}

There are different ways to describe an elliptic curve, the simplest being as a hypersurface or, more generally, as a complete intersection of some weighted projective space.
As an example consider the weighted projective space ${\mathbb P}_{2,3,1}$ spanned by the homogeneous coordinates $(x,y,z) \simeq (\lambda^2 x, \lambda^3 y, \lambda z)$, where $(x,y,z)$ are complex coordinates and $\lambda \subset {\mathbb C}^*$. Due to the scaling relation this space is 2-complex dimensional. An elliptic curve is a flat 1-complex dimensional space, i.e.\ a Calabi-Yau 1-fold. It can thus be described by the hypersurface cut out by the vanishing locus of a homogenous polynomial in $(x,y,z)$ of degree $6$ under rescaling by $\lambda$. Here we used the general fact that a hypersurface in weighted projective space is Ricci-flat whenever the degree of its defining polynomial equals the sum of degrees of the homogenous coordinates.  
A degree-six polynomial in ${\mathbb P}_{2,3,1}$ defines the space ${\mathbb P}_{2,3,1}[6]$. One can show that after suitable coordinate redefinitions such a polynomial can always be brought into the so-called Weierstrass form
\bea
\label{Sec-NP-Weiersrass1}
P_{\text W}= y^2 - x^3 - f x z^4 - g z^6 = 0,
\eea
where $f, g \in \mathbb C$ specify the shape of the elliptic curve as detailed below.
Note at this stage that alternative representations of elliptic curves along these lines are the hypersurfaces  $\mathbb P_{1,1,2}[4]$ or $\mathbb P_{1,1,1}[3]$.

As argued at the end of section \ref{need}  the physics of the 7-branes is encoded in the degenerations of the elliptic fiber.
A general mathematical fact is that a hypersurface described by the equation $P_{\rm W}=0$ becomes singular whenever
\bea
P_{\text W}=0     \quad\quad  {\rm and} \quad\quad      dP_{\text W}=0,
\eea
where the last condition indicates a degeneration of the tangent space.
It is easy to work out these conditions for the Weierstrass curve (\ref{Sec-NP-Weiersrass1}).
First use the scaling relations to go to inhomogeneous coordinates by setting $z=1$.
Then the above conditions are equivalent to
\bea
\label{roots}
&& y=0, \nonumber \\
&&  - x^3 - f x  - g  = (x- a_1)  (x- a_2)   (x- a_3) = 0,  \\
&&  (x- a_1)  (x- a_2)  +  (x- a_2)  (x- a_3)   +  (x- a_1)  (x- a_3) =0,  \nonumber 
\eea
where $a_1, a_2, a_3$ are the three complex roots of the cubic polynomial $ - x^3 - f x  - g $.
Clearly the above requires that two or more of these roots coincide. 
As always, the structure of the roots of this cubic polynomial is encoded in the  discriminant $\Delta$, which vanishes if and only if at least two roots coincide.
For the Weierstrass model the discriminant of the cubic  $- x^3 - f x  - g$ takes the simple form
\bea
\label{Sec-NP-Delta}
\Delta = 27 g^2 + 4 f^3,
\eea 
which shows that the structure of the elliptic curve indeed depends on the parameters $f$ and $g$.

In fact, the dependence of the complex structure $\tau$ on $f$ and $g$ can be made more precise.
Since $\tau$ undergoes $SL(2,\mathbb Z)$ monodromies it is more convenient to work with the $SL(2, \mathbb Z)$ invariant Jacobi $j$-function, which yields an isomorphism from the fundamental domain of $SL(2, \mathbb Z)$ to the Riemann sphere. Instead of introducing the formal definition of $j(z)$ in terms of theta-functions it suffices here to note the expansion
\bea
\label{jf}
j (z) = {\rm exp}(-2 \pi i z) + 744 + \mathbb({\rm exp}(2 \pi i z)) + \ldots.
\eea
A classical mathematical result is  that the complex structure $\tau$ of a Weierstrass elliptic curve is determined by $f,g$ via the relation
\bea
\label{Sec-NP-J}
j(\tau)= \frac{4 (24 f)^3}{\Delta}.
\eea

\subsubsection*{From elliptic curves to elliptic fibrations}

Armed with this representation of a single elliptic curve, we can proceed to elliptic fibrations.
Suppose we have some $n$-complex dimensional manifold $B_n$, covered by local coordinates $u_i$. Then a fibration of the Weierstrass curve (\ref{Sec-NP-Weiersrass1}) over $B_n$ is obtained by promoting the constants $f$ and $g$ in (\ref{Sec-NP-Weiersrass1}) to suitable polynomials in the  coordiates $u_i$ of $B_n$, $f = f(u_i)$,   $g= g(u_i)$. This defines the corresponding Weierstrass model, also known as an $E_8$-fibration. Note that in principle one can also consider fibrations of the elliptic curves $\mathbb P_{1,1,2}[4]$ or $\mathbb P_{1,1,1}[3]$ (called $E_7$- and, respectively, $E_6$-fibrations) and their generalisations. The crucial point, however, is the following mathematical theorem (see e.g.\ \cite{Deligne}):  Every elliptic fibration with a section\footnote{See \cite{Witten:1996bn} for a discussion of F-theory on elliptic fibrations \emph{without} such a section.} can be represented by a Weierstrass model defined in terms of the equation (\ref{Sec-NP-Weiersrass1}) with varying $f$ and $g$.  All other representations such as the $E_7$- and $E_6$-fibrations are birationally equivalent (equivalent up to a flop transition).

Via the relation (\ref{Sec-NP-J}) the complex structure $\tau$ is now dependent on the base coordinates $u_i$. In particular, the elliptic fiber degenerates on a codimension-one sublocus on $B_n$ determined by the vanishing of the likewise $u_i$-dependent discriminant $\Delta= 27 g^2 + 4 f^3$. In view of what we said before, this vanishing locus must be interpreted as a divisor wrapped by a stack of 7-branes.
Indeed, suppose $\Delta$ vanishes to order $N$ along the divisor $S$ on $B_n$  defined by $w=0$. Then eqns. (\ref{jf}) and (\ref{Sec-NP-J}) imply that in the directions normal to this divisor
\bea
\tau \simeq \frac{N}{2\pi i} \, {\rm ln}(w)
\eea
up to regular terms, which is exactly the behaviour expected for $N$ coincident 7-branes. 
The connection between this geometric description of 7-branes as the degeneration locus of the elliptic fibration and the concept of $[p,q]$-7-branes comes about as follows: A single $[p,q]$-7-brane is the locus on the base $B_n$ along which a $(p,q)$-cycle on the elliptic fiber degenerates. This can be shown by matching the geometric $SL(2,\mathbb Z)$ monodromies with the action (\ref{Mpq}) of $M_{p,q}$. Locally around each 7-brane we are free to declare the degenerate 1-cycle in the fiber to correspond to the $(1,0)$ fiber by a choice of basis for $H_2(T^2,\mathbb Z)$. This corresponds to the assertion that locally each brane looks like a D7-brane by an $SL(2,\mathbb Z)$ transformation.
 In IIB theory, a stack of $N$ coincident 7-branes in generic position with respect to the O7-plane gives rise to gauge group $U(N)$. In the F-theory picture the gauge group along a divisor is described by the details of the degeneration of the elliptic fiber as we will see below.

First, however, we need to discuss the Calabi-Yau property in greater detail.
The first Chern class of an elliptic fibration $Y_{n+1}$ is related to the first Chern class of the base space $B_n$ and the degeneration locus of the fibration as  \cite{Kodaira}
\bea
c_1(T_{Y_{n+1}}) \simeq  \pi^* \Bigl(  c_1(T_{B_{n}}) - \sum_i \frac{\delta_i}{12}   [\Gamma_i] \Bigr), \quad\quad\quad\quad \delta_i = {\cal O} (\Delta)|_{\Gamma_i}.
\eea
Here
\bea
\pi: Y_{n+1} \rightarrow B_n
\eea
denotes the projection from the fibration to the base, and its pullback $\pi^*$ maps $H^2(B_{n}, \mathbb Z) \rightarrow H^2(Y_{n+1}, \mathbb Z)$.
Furthermore, the discriminant is supposed to vanish to order $\delta_i$ along the divisors $\Gamma_i$ on the base, with dual 2-form $[\Gamma_i]$.
Strictly speaking this formula is correct only for elliptic $K3$-manifolds, but the extra complications due to higher codimension degenerations on the base do not affect the present argument \cite{Morrison:1996na}. 
We see that the Calabi-Yau property of $Y_{n+1}$ forces $B_n$ to acquire positive curvature such as to compensate for the degenerations of the fiber.
In particular, $B_{n}$ itself is not Calabi-Yau. Since the discriminant locus describes the positions of the 7-branes, we find that the total amount of 7-branes leading to a consistent compactification is thus constrained by the curvature of the base space via
\bea
\label{Sec-NP-12c1}
\sum_i {\delta_i}  \,  [\Gamma_i] = 12 \,  c_1(T_{B_{n}})  .
\eea
This equation is reminiscent of the 7-brane tadpole cancellation condition in perturbative Type IIB theory
\bea
\sum_i N_i [\Gamma_i] = 4 [O_7] .
\eea
This demonstrates the very different nature of string model building in perturbative Type IIB  orientifolds and F-theory in that the 7-brane consistency conditions are \emph{automatically} incorporated in a well-defined geometry.

From a mathematical perspective, the relation (\ref{Sec-NP-12c1}) constrains the degree of the polynomials $f$ and $g$ that appear in the Weierstrass model as follows:
The class of the left-hand side is just the class of the full discriminant $\Delta$, and  (\ref{Sec-NP-12c1}) tells us that $[\Delta] = 12 c_1(T_{B_n})$.
In view of $\Delta = 27 g^2 + 4 f^3$ this determines the class of $f, g$. In general $f$ and $g$ are not globally defined functions of the base space, but rather sections of a line bundle ${\cal L}$. In terms of the canonical bundle $K_{B_n}$, which has first Chern class $c_1({K_{B_n}}) = -c_1(T_{B_n})$, it follows that $f$ and $g$ are sections of $K_{B_n}^{-4}$ and $K_{B_n}^{-6}$, respectively.
Homogeneity of the defining Weierstrass polynomial then forces also $x$ and $y$ to transform as sections of the base. In all, we find
\bea
&& x \in H^0(B_n, K_{B_n}^{-2}), \quad\quad y \in H^0(B_n, K_{B_n}^{-3}), \quad\quad z \in H^0(B_n, {\cal O}), \\
&& f \in H^0(B_n, K_{B_n}^{-4}), \quad\quad g \in H^0(B_n, K_{B_n}^{-6}).
\eea

\subsubsection*{Example: F-theory on $K3$}

As a simple example consider $K3$ on the locus of its moduli space where it arises as an elliptic fibration over $B_1=\mathbb P^1$.  In terms of the normalised volume form $t$ of $\mathbb P^1$, the first Chern class of $K_{{\mathbb P}^1}$ is simply $c_1(K_{{\mathbb P}^1}) = -2 t$. This follows from the Hirzebruch-Riemann-Roch theorem, whereby 
\bea
\chi = 2 -2g = \int_{{\mathbb P}^1} c_1(T_{{\mathbb P}^1}).
\eea
The standard notation is of course $K_{{\mathbb P}^1} = {\cal O}(-2)$.
More generally a section of the line bundle ${\cal O}(-n)$ is a homogenous polynomial of degree $n$ in the homogeneous coordinates $[u_0,u_1]$ of $\mathbb P^1$.
Together with the original scaling relation of the Weierstrass model the elliptically fibered $K3$ is spanned by the coordinates
\bea
(u_0,u_1; x,y,z) \simeq  (u_0,u_1; \lambda^2 x,\lambda^3 y, \lambda z) \simeq  (\mu u_0,\mu u_1; \mu^4 x,\mu^6 y, z) .
\eea
Famously, the discriminant $\Delta$ is thus a polynomial of degree $24$ on the base, with $24$ zeroes known as the position of the 24 7-branes of compactification of IIB theory on ${\mathbb P}^1$. Not all of these 24 branes describe perturbative D7-branes. As discussed already around eqn.~(\ref{Oaction}), in the strict perturbative limit the tadpoles are cancelled locally by grouping the 24 branes into four stacks of six branes such that each stack corresponds to 4 D7-branes, i.e.\ $A$-type branes, on top of a system of $BC$ branes. Recall that the latter furnish the F-theoretic description of the IIB O7-plane.

At a more advanced level, the F-theory solution teaches us several exciting lessons: In \cite{Sen:1996vd} Sen showed how the O7-brane splits as one removes one or more A-branes from the $A^4BC$ system. Recall from the discussion after (\ref{NP-Equ-Gauss}) that in such a situation where tadpoles are no longer cancelled locally the naive supergravity solution must break down close to the O7-plane. In F-theory, the geometry adjusts itself via this non-perturbative brane split so as to render the configuration self-consistent.
This not only reveals that the O7-plane is a dynamical object per se; it also furnishes a highly non-trivial check that F-theory correctly captures the non-perturbative degrees of the theory. 
Indeed it is quite remarkable that a comparatively  straightforward analysis of the geometry of $K3$ alone dictates all consistent configurations $[p,q]$-brane configurations. Along these lines, the authors of \cite{Dasgupta:1996ij} found more general solutions with constant, but non-perturbatively large axio-dilaton for which the 24 branes group e.g.\ into three bunches of the type $A^5 B C^2$ and others. This illustrates that the number of branes of a given $[p,q]$-type on a given compactification space changes from configuration to configuration due to the appearance of monodromies as we start moving the branes around.

\subsection{Sen's orientifold limit }
\label{sec_Sen}

After this first encounter with the geometric description of F-theory, let us go back and analyse more generally how to recover the weakly coupled Type IIB orientifold picture. This is accomplished by a procedure due to Sen \cite{Sen:1997gv}.
Recall from the discussion after eq.~(\ref{NP-Equ-Gauss}) that in the IIB limit one considers the 7-branes as probe objects and takes the axio-dilaton as non-varying and perturbatively small. We already noted that strictly speaking this can only be realised if all 7-brane tadpoles are cancelled locally by placing a suitable amount of 7-branes on top of the orientifold planes.
More generally, there can however exist a limit in which one can \emph{approximately} take $\tau$ small and non-varying even though not all tadpoles are cancelled locally. 
As argued after eqn. (\ref{lambda}) this corresponds to the limit $\lambda \rightarrow \infty$.
Since the profile of $\tau$ is encoded in the Jacobi function (\ref{Sec-NP-J}) a constant $\tau$ can be achieved as long as $j(\tau)$ is non-varying over the base $B_3$ of the F-theory 4-fold. To arrange for this one makes
the general ansatz 
\bea
f= - 3 h^2 + \epsilon \eta, \quad\quad\quad g= - 2 h^3  + \epsilon h \eta  - \frac{\epsilon^2}{12} \chi,
\eea
where $\epsilon$ is an arbitrary constant. Sen's orientifold limit corresponds to the limit $\epsilon \rightarrow 0$ with $h$, $\eta$ and $\chi$ generically non-vanishing.
In this case one finds that indeed the string coupling becomes arbitrarily weak everywhere except at the location $h=0$, where $g_s \rightarrow \infty$. It can thus be considered as a perturpative parameter of the theory.
The discriminant locus factorises to leading order in $\epsilon$ as
\bea \label{Deltae}
  \Delta_{\epsilon} = - 9 \epsilon^2 h^2 (\eta^2 - h \chi) + {\cal O}(\epsilon^3) 
\eea
so that the fibration degenerates at $h=0$ and at $\eta^2 - h \chi$.  
The latter may factorise further depending on the particular form of the polynomials $\eta$ and $\chi$.
By closer inspection of the monodromies around these loci it can be shown that $h=0$ corresponds to the location of an O7-plane, while the latter locus is that of an ordinary D7-brane. 
The inclusion of higher order terms in $\epsilon$, on the other hand, destroys this factorisation of $\Delta$ into orientifold and brane piece. This process can be thought of as recombination of the O7-plane and the D7-brane. Since in perturbation theory there are no light strings stretching between the O7-plane and a D7-brane, this  recombination is truly non-perturbative. The involved recombination modes are expected to be  $\tiny{{\begin{pmatrix} p   \\  q   \end{pmatrix}}}$ strings.

The Type IIB theory is defined on the Calabi-Yau 3-fold $X$ which is a double cover of the base $B$ branched over $h=0$. 
Concretely, suppose the base $B_3$ is embedded into some projective space with projective coordinates $u_i$. Then one can define $X$ as the complete intersection in the projective space spanned by the coordinates $u_i$ and $\xi$ which is defined by the original hypersurface polynomial and the additional constraint
\bea
\label{xi2}
\xi^2  -h(u_i) =0 .
\eea
Conversely, given a Type IIB orientifold on $X$ together with a $\mathbb Z_2$ action $\sigma$, one can construct the F-theory uplift by reversing Sen's limit above. This has been extended and applied in the more recent literature \cite{Collinucci:2008zs,Blumenhagen:2009up,Collinucci:2009uh}.
 
Note from the form of the 7-brane locus $\eta^2 - \xi^2 \chi=0$ (where we replaced $h$ in $\eta^2 - h  \chi=0$ by $\xi^2$ according to (\ref{xi2}))  that for the defining equation for a single D7-brane configuration with a well-defined F-theory uplift is non-generic \cite{Braun:2008ua,Collinucci:2008pf}. Brane-image brane configurations require a specialisation of the polynomials $\eta$ and $\chi$, e.g.\ by  factorisation of $\chi$. 
The relation of such even more non-generic brane splits to elliptic fibrations of $E_7$- and $E_6$-type has been investigated in \cite{Aluffi:2009tm,GW-new}.

Finally we would like to point out that the limit $\epsilon \rightarrow 0$ with all other polynomials generic is not the only way to make $\tau$ non-varying along $B_3$.
There exist in addition the two possibilities of taking either $f \equiv 0$ or $g \equiv 0$ \cite{Dasgupta:1996ij}. In these cases, $\tau$ is constant and fixed at non-perturbatively large values, in agreement with the appearance of exceptional gauge groups. These configurations correspond to generalisations of the $\mathbb Z_2$ orientifold action of the strongly coupled Type IIB string to $\mathbb Z_n$ actions.

\subsection{Gauge symmetry from degenerations}
\label{sec_Gau}

From a physics perspective the most essential data of an elliptic fibration are the locus and the type of fiber degenerations because these allow us to deduce the nature of the 7-branes wrapping the corresponding divisor.
The different ways how the complex structure of the elliptic fiber can degenerate have been classified by Kodaira \cite{Kodaira} for the case of a Weierstrass model of $K3$. The relevant criterion to be imposed here is that the  singularities can be resolved without destroying the Calabi-Yau property.
These results have subsequently been generalised to higher Calabi-Yau spaces (see \cite{Bershadsky:1996nh}\footnote{\label{foot} Note that this reference is devoted to an analysis of elliptic Calabi-Yau 3-folds. For higher dimensional manifolds additional effects due to a more complicated monodromy structure can occur, and as of this writing no complete classification exists in the literature. Also, the results of  ref.~\cite{Bershadsky:1996nh} apply to situations where the discriminant locus itself is non-singular as a divisor of $Y_3$.}
 and references therein for a physics discussion).

Consider a Weierstrass model  $Y_{n+1}$. One distinguishes degenerations that merely render the fibration singular without inducing an actual singularity in $Y_{n+1}$ versus loci where the Calabi-Yau $(n+1)$-fold becomes singular itself.
The simplest kind of degeneration of the first type corresponds to a so-called $I_1$ singularity, which arises in the fiber over just a single 7-brane. For a $[p,q]$-7-brane the $(p,q)$ 1-cycle of the elliptic fiber pinches off and the elliptic fiber forms a sausage-type $\mathbb P^1$ whose north and south pole touch each other. 
Following Kodaira, $I_1$ singularities arise in a Weierstrass model whenever $\Delta$ develops a simple zero while $f$ and $g$ are non-zero. In this case two of the roots $a_i$ in the cubic (\ref{roots}) coincide. From eqs.\ (\ref{jf}), (\ref{Sec-NP-J}) the simple zero in $\Delta$ leads to the logarithmic profile (\ref{NP-Equ-Gauss}) of $\tau$  characteristic of a single 7-brane.

When $\Delta$ vanishes to higher order,
generically also $Y_{n+1}$ develops an actual singularity as a manifold.  For the simplest case of $Y_2  = K_3$ the possible singularities that can occur are precisely the ones listed in the famous ADE classification \cite{Kodaira}. For higher dimensional $Y_{n+1}$ with $n > 2$ also the action of monodromies along the brane have to be taken into account. This can create more general singularities including non-simply laced examples of $B$ and $C$ type and such exceptional cases as $G_2$ or $F_4$.

Let us focus for definiteness on Calabi-Yau 4-folds $Y_4$.
One intuitive way to understand the connection between a singularity over the divisor $S \subset B_3$ and a Lie algebra $G$ is as follows: The singularity in $Y_{4}$ can be resolved by standard methods of algebraic geometry. 
For our purposes it suffices to consider cases that allow for a so-called split simultaneous resolution.
As described e.g. in \cite{Katz:1996th} this means that there exists a new, non-singular Calabi-Yau $\ov Y_{4}$ defined by replacing the singular fiber over $S$ by a tree of $\mathbb P^1$s  which we call in the sequel $\Gamma_i$, $i = 1, \ldots, {\rm rk}(G)$. Note that in this process $h^{1,1}$ increases by ${\rm rk}(G)$. These $\mathbb P^1$s have the property that they intersect one another like the simple roots of the Lie algebra $G$ in the following sense:
The resolution 2-cycles $\Gamma_i$ are fibered over the 7-brane $S \subset B_3$. Denote the resulting divisors of $\ov Y_4$ by 
\bea
\label{res}
\hat D_i: \Gamma_i \rightarrow S.
\eea 
By construction these are $\mathbb P^1$-fibrations. Consider furthermore the linear combination
$\hat D_0 = \hat S - \sum_i a_i D_i$, where $\hat S$ is the elliptic fibration over $S$ and $a_i$ denote the Dynkin labels for the Lie algebra $G$ associated with the singularity over $S$. Then $\hat D_0, \hat D_i$ encode the extended Dynkin diagram of $G$ in the sense that their Poincar\'e dual 2-forms $[\hat D_i] \in H^2(\ov Y_4)$ satisfy
\bea
\label{Cij}
\int_{\ov Y_4} [\hat D_i] \wedge [\hat D_j] \wedge \tilde \omega = - C_{ij} \int_S  \tilde \omega, \quad\quad i,j,= 0, 1, \ldots, {\rm rk}(G)
\eea
for  $\tilde \omega \in H^2(B_3)$ and $C_{ij}$ the Cartan matrix of $G$. Conversely, the singular F-theory limit can be understood as the limit in which the $\Gamma_i$ shrink to zero volume. The singularity is the result of a collision of ${\rm rk}(G)$ zero size ${\mathbb P^1}$s.

The appearance of the corresponding gauge groups along the 7-brane wrapping the singular locus can be understood by F/M-theory duality.
As sketched in (\ref{C3}), Kaluza-Klein reduction of $C_3$ along non-singular fibers results in the Type IIB closed string 2-forms $B_2$ and $C_2$. 
As we have just seen, in the fiber above the discriminant locus extra two-cycles appear along which $C_3$ can be reduced.
For simplicity consider an $A_{n-1}$ singularity associated with gauge group $SU(n)$. The resolution $\mathbb P^1$s $\Gamma_i, i=1, \ldots (n-1)$,  correspond to the nodes of the Dynkin diagram of $A_{n-1}$. Massless vector states arise from two sources:
\begin{itemize}
\item The M-theory 3-form $C_3$ can be reduced along $\Gamma_i$, leading to 1-forms $A_i = \int_{\Gamma_i} C_3$ along the 7-brane. These are interpreted as the gauge potentials for the abelian group factors in the Cartan subalgebra of $SU(n)$.
\item The M-theory M2-brane can wrap chains of 2-cycles $S_{ij} = \Gamma_i \cup  \Gamma_{i+1} \cup \ldots \cup \Gamma_j$ for any $i \leq j$. Taking into account the two possible orientations this gives rise to $n^2 - n$ states along the brane. In the singular limit of vanishing volume of the $\Gamma_i$ these states become massless and lead to the $W$-bosons of $SU(n)$.
\end{itemize}
In all this yields the $n^2 -1$ generators of $SU(n)$.  This logic can be generalised to other simple groups.

In addition there can be extra abelian gauge factors which do not arise as the Cartan generators of a non-abelian gauge group and which are strictly speaking not localised along individual divisors.
An unambiguous way to determine the total rank of the gauge group of a singular Weierstrass model is to note that all $U(1)$ generators derive from reduction of $C_3$ as $C_3 = A_i \wedge \omega_i + \ldots$, where the two-forms $\omega_i$ have one leg along the $T^2$ fiber and one leg along the base \cite{Morrison:1996na,Morrison:1996pp}.
Therefore 
\bea
r = h^{1,1}(\ov Y_{n+1}) -  h^{1,1}(B_n) - 1
\eea
gives the total rank of the F-theory gauge group including possible abelian gauge factors.\footnote{In addition, there are $h^{2,1}(B_n)$ bulk $U(1)$ fields, corresponding to reduction of the RR 4-form $C_4$ along 3-cycles on $B_n$.} The physics of these extra $U(1)$s in compactifications to four dimensions has been discussed in \cite{GW-new}.

Let us stress that for model building purposes it is essential to have control over the resolution $\ov Y_4$ of the singular fibration $Y_4$, e.g. to determine the complete rank of the gauge group or in the context of 3-brane tadpole cancellation, see section \ref{sec_Flu}. In fact, for the class of Calabi-Yau 4-folds constructed as hypersurfaces or complete intersections of toric spaces powerful tools have been developed to perform this resolution in terms of the divisors $\hat D_i$ explicitly \cite{Candelas:1996su,Candelas:1997eh}. These have recently been exploited in the context of F-theory model building in \cite{Blumenhagen:2009yv,Krause,Chen:2010ts,GW-new}.

To conclude this section, we point out that the appearance of exceptional and more general gauge groups has a beautiful interpretation in terms of  $\tiny{{\begin{pmatrix} p   \\  q   \end{pmatrix}}}$-strings \cite{Gaberdiel:1997ud,DeWolfe:1998zf}. This can be understood already for F-theory on $K3$, where the possible gauge groups are exhausted by the ADE classification. By definition, only  $\tiny{{\begin{pmatrix} p   \\  q   \end{pmatrix}}}$-strings can end on $[p,q]$-7-branes. Due to the appearance of monodromies, however, $\tiny{{\begin{pmatrix} p   \\  q   \end{pmatrix}}}$-strings stretched between two branes along different paths can lead to inequivalent states. A mild version of this phenomenon was encountered already in the perturbative limit. A  $\tiny{{\begin{pmatrix} 1   \\  0   \end{pmatrix}}}$-string encircling a $BC$-system gets orientation reversed, see eqn.\ (\ref{Oaction}). The resulting  unoriented strings correspond to strings between a stack of branes and their orientifold images. More generally, one can classify allowed paths that lead to BPS-strings and determine their interactions via consistent splitting and joining. Suffice it here to highlight from  \cite{Gaberdiel:1997ud} that the various $\tiny{{\begin{pmatrix} p   \\  q   \end{pmatrix}}}$-strings occurring for the $[p,q]$-7-brane system $A^n B C^2$ indeed span the adjoint of $E_{5+n}$, $n=1,2,3$, in agreement with the monodromies for the generalised orientifolds of \cite{Dasgupta:1996ij}. The underlying reason for this new richness is that $\tiny{{\begin{pmatrix} p   \\  q   \end{pmatrix}}}$-strings of different types can form string junctions with more than just two endpoints. This overcomes the limitation to  two-index representations as carried by the Chan-Paton factors of perturbative $\tiny{{\begin{pmatrix} 1   \\  0   \end{pmatrix}}}$-strings. The latter allow only for the construction of $U(N)$, $Sp(2N)$ and $SO(N)$ groups.

\section{Technology for F-theory compactifications}
\label{Technology}

 \subsection{Tate models}
 \label{sec_Tat}
 
Given the importance of the singularity structure for an analysis of the F-theory landscape, we find it useful to present the methods to detect the singularities in slightly greater detail. There exists an algorithm due to Tate that allows one to  read off the singularity structure of an elliptic fibration, say  $Y_4: T^2 \rightarrow B_3$. 
Tate's formalism consists in a local coordinate redefinition that brings the Weierstrass constraint
$P_{\rm W}=0$ in (\ref{Sec-NP-Weiersrass1}) into the Tate form 
\bea 
\label{Tate1}
    P_{\rm W} = x^3 - y^2 + x\, y\,  z\, a_1 + x^2\, z^2\, a_2 + y\, z^3\,a_3   + x\, z^4\, a_4 + \, z^6\, a_6\ = 0.
\eea
For a general Weierstrass model this can be achieved \emph{locally} via the so-called Tate algorithm described in detail in~\cite{Bershadsky:1996nh}. 
In the sequel we will only be working with the inhomogeneous Tate form by setting $z=1$. The $a_n(u_i)$ depend on the complex coordinates $u_i$ of the base $B$.
 They encode the discriminant locus of the elliptic fibration. To recover the Weierstrass model one first introduces the combinations
\beq
  \beta_2 = a_1^2 + 4 a_2 ,\qquad 
  \beta_4 = a_1 a_3 + 2\, a_4 ,\qquad
  \beta_6 = a_3^2 + 4 a_6.
\eeq
One can check by completing the square and the cube in  (\ref{Tate1}) that the Weierstrass sections $f$ and $g$ are related to these via
\bea
\label{fg}
f=-\frac{1}{48}( \beta_2^2 -24\, \beta_4), \qquad
  g=-\frac{1}{864}( -\beta_2^3 + 36 \beta_2 b_4 -216 \, \beta_6) .
\eea
The discriminant $\Delta = 27 g^2 + 4 f^3$ can then be expressed as
\beq
  \Delta = -\tfrac14 \beta_2^2 (\beta_2 \beta_6 - \beta_4^2) - 8 \beta_4^3 - 27 \beta_6^2 + 9 \beta_2 \beta_4 \beta_6.
\eeq
In general, the discriminant $\Delta$ will factorize with each factor describing the location of a 7-brane on a divisor in $B_3$. The precise group is encoded by the vanishing degree of the $a_i$ and $\Delta$.
This has been classified in Table~2 of ref.~\cite{Bershadsky:1996nh}, which we are reproducing (in the form given in \cite{Blumenhagen:2009up}) for convenience of the reader in table \ref{tab:TateTable}.\footnote{Note that this table was derived for the case of elliptic Calabi-Yau 3-folds, see footnote \ref{foot}.}

\label{appb}
\begin{table}[h!]
        \centering
                \begin{tabular}{c|c|cc|cc@{${}\,\,\,\quad$}ccc}
                        sing.                        & discr.         & \multicolumn{2}{|c}{gauge enhancement}        & \multicolumn{5}{|c}{coefficient vanishing degrees}  \\
                        type                         & $\deg(\Delta)$ & type & group  & ${}\,\,\,\,\,\,a_1\quad{}$ & $a_2$ & $a_3$ & $a_4$ & $a_6{}_\big.$ \\ \hline\hline
                        $\mathrm{I}_0{}^\big.$               & 0      &            & ---          & 0 & 0 & 0     & 0     & 0 \\
                        $\mathrm{I}_1$                       & 1      &            & ---          & 0 & 0 & 1     & 1     & 1 \\
                        $\mathrm{I}_2$                       & 2      & $A_1$      & $SU(2)$      & 0 & 0 & 1     & 1     & 2 \\
                        $\mathrm{I}_{2k}^{\,\mathrm{ns}}$    & $2k$   & $C_{2k}$   & $SP(2k)$     & 0 & 0 & $k$   & $k$   & $2k$${}^\big.$ \\
                        $\mathrm{I}_{2k}^{\,\mathrm{s}}$     & $2k$   & $A_{2k-1}$ & $SU(2k)$     & 0 & 1 & $k$   & $k$   & $2k$ \\
                        $\mathrm{I}_{2k+1}^{\,\mathrm{ns}}$  & $2k+1$ &            & [unconv.]    & 0 & 0 & $k+1$ & $k+1$ & $2k+1$ \\
                        $\mathrm{I}_{2k+1}^{\,\mathrm{s}}$   & $2k+1$ & $A_{2k}$   & $SU(2k+1)$   & 0 & 1 & $k$   & $k+1$ & $2k+1$${}_\big.$ \\ \hline
                        $\mathrm{II}$                        & 2      &            & ---          & 1 & 1 & 1     & 1     & 1${}^\big.$ \\
                        $\mathrm{III}$                       & 3      & $A_1$      & $SU(2)$      & 1 & 1 & 1     & 1     & 2 \\
                        $\mathrm{IV}^{\,\mathrm{ns}}$        & 4      &            & [unconv.]    & 1 & 1 & 1     & 2     & 2 \\
                        $\mathrm{IV}^{\,\mathrm{s}}$         & 4      & $A_2$      & $SU(3)$      & 1 & 1 & 1     & 2     & 3 \\
                        $\mathrm{I}_0^{*\,\mathrm{ns}}$      & 6      & $G_2$      & $G_2$        & 1 & 1 & 2     & 2     & 3 \\
                        $\mathrm{I}_0^{*\,\mathrm{ss}}$      & 6      & $B_3$      & $SO(7)$      & 1 & 1 & 2     & 2     & 4 \\
                        $\mathrm{I}_0^{*\,\mathrm{s}}$       & 6      & $D_4$      & $SO(8)$      & 1 & 1 & 2     & 2     & 4 \\
                        $\mathrm{I}_1^{*\,\mathrm{ns}}$      & 7      & $B_4$      & $SO(9)$      & 1 & 1 & 2     & 3     & 4 \\
                        $\mathrm{I}_1^{*\,\mathrm{s}}$       & 7      & $D_5$      & $SO(10)$     & 1 & 1 & 2     & 3     & 5 \\
                        $\mathrm{I}_2^{*\,\mathrm{ns}}$      & 8      & $B_5$      & $SO(11)$     & 1 & 1 & 3     & 3     & 5 \\
                        $\mathrm{I}_2^{*\,\mathrm{s}}$       & 8      & $D_6$      & $SO(12)$     & 1 & 1 & 3     & 3     & 5${}_\big.$ \\ \hline
                        $\mathrm{I}_{2k-3}^{*\,\mathrm{ns}}$ & $2k+3$ & $B_{2k}$   & $SO(4k+1)$   & 1 & 1 & $k$   & $k+1$ & $2k$${}^\big.$ \\
                        $\mathrm{I}_{2k-3}^{*\,\mathrm{s}}$  & $2k+3$ & $D_{2k+1}$ & $SO(4k+2)$   & 1 & 1 & $k$   & $k+1$ & $2k+1$ \\
                        $\mathrm{I}_{2k-2}^{*\,\mathrm{ns}}$ & $2k+4$ & $B_{2k+1}$ & $SO(4k+3)$   & 1 & 1 & $k+1$ & $k+1$ & $2k+1$ \\
                        $\mathrm{I}_{2k-2}^{*\,\mathrm{s}}$  & $2k+4$ & $D_{2k+2}$ & $SO(4k+4)$   & 1 & 1 & $k+1$ & $k+1$ & $2k+1$${}_\big.$ \\ \hline
                        $\mathrm{IV}^{*\,\mathrm{ns}}$       & 8      & $F_4$      & $F_4$        & 1 & 2 & 2     & 3     & 4${}^\big.$ \\
                        $\mathrm{IV}^{*\,\mathrm{s}}$        & 8      & $E_6$      & $E_6$        & 1 & 2 & 2     & 3     & 5 \\
                        $\mathrm{III}^*$                     & 9      & $E_7$      & $E_7$        & 1 & 2 & 3     & 3     & 5 \\
                        $\mathrm{II}^*$                      & 10     & $E_8$      & $E_8$        & 1 & 2 & 3     & 4     & 5 \\
                        non-min                              & 12     &            & ---          & 1 & 2 & 3     & 4     & 6
                \end{tabular}
        \caption{Refined Kodaira classification resulting from Tate's algorithm. In order to distinguish the ``semi-split'' case $\mathrm{I}_{2k}^{*\,\mathrm{ss}}$ from the ``split'' case $\mathrm{I}_{2k}^{*\,\mathrm{s}}$ one has to work out a further factorization condition, see \S 3.1 of \cite{Bershadsky:1996nh}.} 
                \label{tab:TateTable}
\end{table}

 For example, an $SU(5)$ gauge group along the divisor 
 \bea
 S: \quad w=0 
 \eea
 corresponds to
\beq \label{TateSU5b1}
  a_1 = \mathfrak{b}_5 , \quad
  a_2 = \mathfrak{b}_4 w , \quad
  a_3 = \mathfrak{b}_3 w^2 , \quad
  a_4 = \mathfrak{b}_2 w^3 , \quad
  a_6 = \mathfrak{b}_0 w^5 ,
  \eeq
where the $\mathfrak{b}_i$ generically depend on all coordinates $u_i$ of the base $B$ but do not contain an overall factor of $w$. 
It is straightforward to evaluate the discriminant $\Delta$ of the elliptic fibration in terms of the new sections $\mathfrak{b}_i$ as
\beq \label{Delta_SU5}
  \Delta =  - w^5 \, \left(  \mathfrak{b}_5^4  P  + w  \mathfrak{b}_5^2 (  8  \mathfrak{b}_4 P +  \mathfrak{b}_5  R  ) + w^2 (16  \mathfrak{b}_3^2  \mathfrak{b}_4^2 +  \mathfrak{b}_5 Q) + {\cal O}(w^3) \right)
\eeq
with
\beq
  P = \mathfrak{b}_3^2 \mathfrak{b}_4 - \mathfrak{b}_2 \mathfrak{b}_3 \mathfrak{b}_5 + \mathfrak{b}_0 \mathfrak{b}_5^2 , \qquad
  R = 4 \mathfrak{b}_0 \mathfrak{b}_4 \mathfrak{b}_5- {\mathfrak{b}_3^3} - \mathfrak{b}_2^2 \mathfrak{b}_5 .
\eeq
Generically the expression in brackets in \eqref{Delta_SU5}, denoted as $S_1$, does not factorize further and thus constitutes the single-component locus of an $I_1$ singularity. 
Cohomologically, one thus finds that the class $[\Delta]$ splits as $[\Delta] = 5 [S] + [S_1]$.

As stressed in section \ref{sec_Gen}, the most general elliptic fibration with a section can always be written as a Weierstrass model. A special subclass of such models, however, can be brought into the Tate form not just locally around the discriminant locus, but globally. I.e.\ for those models the defining constraint is (\ref{Tate1}) in terms of globally defined sections $a_i \in H^0(B,K_B^{-i})$. Clearly each such global Tate model defines also a Weierstrass model in that each set of $a_i$ defines the Weierstrass sections $f$ and $g$ as in (\ref{fg}). The converse, however, is not true globally as the transformation from $f$ and $g$ to $a_i$ involves branch cuts.

Models which are globally of the Tate form are very convenient because one can immediately read off the gauge group from table \ref{tab:TateTable} without going through the algorithm of \cite{Bershadsky:1996nh} to bring the Weierstrass polynomial locally into the form (\ref{Tate1}). It is important to keep in mind, though, that they do not give rise to the most general singularity structure. In particular, there is a sense in which such models are based on an underlying gauge group $E_8$. Consider the situation that the gauge group $G$ realised along the divisor $S$ is contained within $E_8$. Then in the global Tate model the dynamics of the gauge sector can be understood via breaking an original $E_8$ symmetry down to the commutant of a complimentary gauge group $H$ such that $E_8 \rightarrow G \times H$. This can be seen by organising the sections of the Tate model as a sum of two terms, the first of which corresponds to singularity enhancement $E_8$, while the second encodes an $H$-bundle responsible for the breaking down to $G$. For more technical information on this point we refer to \cite{GW-new} and references therein.

\subsection{Fluxes and 3-brane tadpoles}
\label{sec_Flu}

While the entire information of 7-brane charge is contained in  the geometric data of the F-theory Calabi-Yau 4-fold,
the induced 5- and 3-brane charge is extra data encoded in the background flux.
In the Type IIB orientifold picture one distinguishes  closed string fluxes  - most notably the three-form flux $G_3=  F_3 - \tau H_3$ - and brane fluxes $F$. The latter are the background value of the Yang-Mills field strength on the 7-branes. 
In F-theory, both bulk 3-form fluxes and brane fluxes enjoy a uniform description in terms of the M-theory four-form flux $G_4$.\footnote{Recently, there has been important progress  \cite{Alim:2009rf,Alim:2009bx,Grimm:2009ef,Aganagic:2009jq,Grimm:2009sy,Jockers:2009ti} in the computation of the $G_4$-flux induced Gukov-Vafa-Witten superpotential of \cite{Gukov:1999ya}.  For background on some of the underlying 4-fold techniques see \cite{Greene:1993vm,Mayr:1996sh,Klemm:1996ts,Lerche:1997zb}; closely related recent work on  B-type brane superpotentials includes  \cite{Jockers:2008pe,Grimm:2008dq,Jockers:2009mn} and references therein.
Various aspects of (flux) potentials in F-theory on $K3 \times K3$ have been analysed in \cite{Gorlich:2004qm,Lust:2005bd, Denef:2005mm, Berglund:2005dm, Aspinwall:2005ad, Braun:2008pz,Valandro:2008zg}.    } In principle, brane fluxes arise by the reduction
\bea
G_4 = \sum_i F^{(i)} \wedge \omega_i + \ldots,
\eea
where $\omega_i$ denote those normalisable harmonic 2-forms of $Y_{n+1}$ that are neither elements of $H^{1,1}(B_n)$ nor represent the fiber class itself. 
For example, this expression straightforwardly describes abelian gauge flux associated with a Cartan $U(1)$ generator contained in the gauge group $G$ along $S$. In this case  $\omega_i \equiv \omega_i^G$  are the 2-forms Poincar\'e dual to the resolution divisors $\hat D_i$ introduced in (\ref{res}). 
Such flux breaks the gauge group $G$ to the commutant of the Cartan generator, a mechanism that will be heavily exploited in the context of GUT symmetry breaking in section \ref{GUTbreaking}. 

More general fluxes, however, are extremely hard to describe in detail because they encode the information of a non-abelian gauge bundle. This is best  understood in the context of the global Tate model introduced before. The flux we have in mind is associated with the orthogonal gauge group $H \subset E_8$ and therefore does not affect the gauge group $G$. An auxiliary tool to study such fluxes is given by the spectral cover construction, which is the subject of section \ref{spectral_cover}. Note that gauge flux is essential to achieve a chiral matter spectrum, see the discussion in the next section.

The total amount of F-theory fluxes is constrained by the 3-brane tadpole cancellation condition, which is dual to 
M2-brane charge cancellation in M-theory. 
Here we simply quote the result \cite{Sethi:1996es}
\bea
\label{Nm2}
\frac{\chi(\ov Y_{4})}{24} = N_{M_2} + \frac{1}{2} \int_{\ov Y_4} G_4 \wedge G_4,
\eea
where $N_{M2}$ is the number of M2-branes filling the three non-compact dimensions. 
By F/M-duality $N_{M2} = N_{D3}$, the number of spacetime-filling D3-branes.

The object ${\chi}(\ov Y_{4})$ is the Euler characteristic, introduced in eqn.\ (\ref{eq-Euler1}), of the resolved Calabi-Yau 4-fold. It encodes all curvature induced 3-form charge of the 7-branes. Indeed in models with a weakly coupled Type IIB description one can match ${\chi}(\ov Y_{4})$ with the curvature-dependent terms in the Chern-Simons action of the O7-plane and D7-branes \cite{Collinucci:2008pf}. 
Special care has to be taken in the computation of $\chi(\ov Y_4)$: As long as the 4-fold is non-singular, $\chi(Y_4) = \int c_4(Y)$, and for a Weierstrass model over $B_3$ a simple formula exists \cite{Sethi:1996es,Klemm:1996ts}
\bea
\label{chismooth}
\chi(Y_4) = 12 \int_{B_3} c_1(B_3) c_2(B_3) + 360 \int_{B_3} c_1^3(B_3) .
\eea
For the phenomenologically relevant case of F-theory compactification with non-abelian gauge groups, however, $Y_4$ is singular and the computation of its Euler characteristic requires the resolution of $Y_4$ to $\ov Y_4$. For general singular Calabi-Yau 4-folds the computation of ${\chi}/{24}$ can become quite involved \cite{Andreas:1999ng,Andreas:2009uf}. However, as stressed above, for the class of 4-folds that can be described with the help of toric methods a well-defined algorithm exists that allows one to deduce ${\chi}/{24}$, see \cite{Blumenhagen:2009yv,Krause,Chen:2010ts} for examples of this type in the context of global F-theory GUT models.
Furthermore for global Tate models with non-abelian singularities solely along a single divisor $S$, a simple closed formula for $\chi/24$ has been conjectured \cite{Blumenhagen:2009yv}, see the discussion around (\ref{chi-prop_app}).

Note that the value of $\chi$ for a singular 4-fold $Y_4$ is usually significantly lower than the expression (\ref{chismooth}) for the corresponding smooth Weierstrass model. This is of direct relevance for the construction of four-dimensional F-theory vacua  because moduli stabilisation and the engineering of a realistic particle spectrum require the inclusion of four-form flux. In consistent models the flux is not allowed to overshoot $\chi/24$ as this would necessitate the inclusion of anti 3-branes, which would lead to instabilities.\footnote{One might argue that the anti 3-brane should be stabilised in a warped region as in KKLT and thus do no further harm, but this assumes a sufficient amount of 3-form flux allowed by the 3-brane tadpole to create a warped throat in the first place.}
Typically, F-theory vacua with high rank non-abelian gauge groups are therefore much more constrained than vacua with lower rank.

\subsection{Matter curves and  Yukawa points}
\label{matter-gen}

In previous sections we have described the localisation of the non-abelian gauge degrees of freedom in F-theory along divisors $D_a$ of the base space $B_{n}$ of the elliptic Calabi-Yau. 
Each such divisor carries a $G_a$ gauge theory with four supercharges, corresponding to ${\cal N}=1$ Super-Yang-Mills in four dimensions.
Besides the vector multiplet in the adjoint representation of $G_a$  there arises extra massless charged matter. 

In Type IIB orientifolds on a Calabi-Yau 3-fold $X$, chiral multiplets in the adjoint representation of $G_a$ correspond to D7-brane moduli \cite{Jockers:2004yj}. These include brane deformation moduli and Wilson line moduli, counted respectively by $H_-^{0,2}(D_a) \simeq H_-^{0}(S, K_S)$ and $H_-^{(0,1)}(S)$. The subscripts indicate that  only the odd eigenspaces of the cohomology groups under the orientifold action  $\sigma$ are relevant. In F-theory there is no clear separation between open and closed string moduli. In fact the brane position moduli become part of the complex structure moduli $H^{(3,1)}(Y_4)$. For a precise account of these fields and the Wilson line moduli in the effective action derived by dimensional reduction of the dual M-theory we refer to \cite{Eff-action}.

Apart from these non-chiral matter fields, there exist two types of potentially chiral charged matter: so-called bulk states that propagate along the whole divisor and matter at the intersection locus of two divisors $D_a$ and $D_b$. The appearance of these chiral multiplets parallels the situation in Type IIB orientifolds, which is therefore worthwhile recalling in this context. For this we will follow mostly the discussion in \cite{Blumenhagen:2008zz}.

\subsubsection*{Bulk matter}

Consider two parallel stacks of Type IIB D7-branes along the same divisor $S$, each carrying internal gauge flux ${2 \pi \alpha' {F}}_a$ and ${2 \pi \alpha' { F}}_b$, respectively. This flux is the curvature of two vector bundles $V_a$ and $V_b$ with Chern character
\bea
 {\rm tr}\left[ \, e^{2\pi\alpha' { F}_i } \right]= {\rm ch}(V_i).
\eea
In the sequel we restrict ourselves to line bundles $L_a$, $L_b$ characterised entirely by their first class $c_1(L_i) =  \frac{1}{2 \pi} {\rm tr} \ell_s^2 F_i$.
For simplicity we assume that $S$ is at general position with respect to the O7-plane so that the gauge group is $U(N_a) \times U(N_b)$.  
The bulk matter describes chiral multiplets in the bifundamental representation
propagating along the whole divisor $S$. Such matter is in general counted by so-called extension groups
\begin{equation}
  \label{ext1} 
  {\rm Ext}^n (\i_* L_a, \i_* L_b), \qquad n=0,\ldots, 3,
  \end{equation}
  where $\i: S \rightarrow X$ denotes the inclusion of the divisor $S$ into the Calabi-Yau 3-fold $X$.
This was derived in detail from an analysis of B-model vertex operators in \cite{Katz:2002gh}.
The value $n=1$ refers to anti-chiral multiplets transforming as $(\ov
N_a, N_b)$, while $n=2$ corresponds to chiral multiplets in the same
representation.  For consistency, the states counted by the groups
corresponding to $n=0$ and $n=3$ must be absent. These states do not
describe matter fields but rather refer to gauge multiplets.
One can show that the sheaf extension
groups eqn.~\eqref{ext1} translate into certain cohomology groups for
the line bundles on the divisor $S$,  concretely
\begin{eqnarray}
  \label{externcoh} 
 { \rm Ext}^0 (\i_* L_a, \i_* L_b)&=& H^0(S, L_a\otimes L^\vee_b), \nonumber \\ 
  {\rm Ext}^1 (\i_* L_a, \i_* L_b)&=& H^1(S, L_a\otimes L^\vee_b)+ 
  H^2(S, L^\vee_a\otimes L_b), \nonumber \\  
  {\rm Ext}^2 (\i_* L_a, \i_* L_b)&=& H^2(S, L_a\otimes L^\vee_b)+ 
  H^1(S, L_a^\vee\otimes L_b), \nonumber \\  
  {\rm Ext}^3 (\i_* L_a, \i_* L_b)&=& \phantom{aaaaaaaaaaaaaaa}\  
  H^0(S, L_a^\vee\otimes L_b). 
\end{eqnarray}
The net chirality follows as \cite{Blumenhagen:2008zz}
\bea
  \label{chiralext} 
    I^{bulk}_{ab} = 
    \sum_{n=0}^3 (-1)^n {\rm dim} \, {\rm Ext}^n (\i_* L_a, \i_* L_b)
   =
    -\int_X [S]\wedge [S]\wedge \left(\, c_1(L_a)-c_1(L_b)\, \right).
\eea
As a variant of this setup one can also consider a single brane stack with gauge group $G$ along $S$ and consider gauge flux in terms of a non-trivial embedding of a vector bundle with structure group $G_2 \subset G$. This breaks $G$ to the commutant $G_1$ of $G_2$.
In this case chiral multiplets charged under $G_1$ arise, which originate from the adjoint representation of $G$.
More precisely, the group theoretic decomposition
\bea
&& G \rightarrow G_1 \times G_2  \\
&& {\rm ad}_{G} \rightarrow ( {\rm ad}_{G_{1}},1)   \oplus (1, {\rm ad}_{G_{2}} ) \oplus \sum (R_x, U_x) 
\eea
leads to bulk matter in suitable representations $R_x$ under the visible gauge group $G_1$.
The individual states are counted by appropriate cohomology groups with values in the bundle representation $U_x$.  For instance,  a non-trivial gauge line bundle $L$ associated with a Cartan $U(1)$ of $G$  breaks $G \rightarrow \tilde G \times U(1)$, and if the representations of $\tilde G$ carry $U(1)$ charge $q$, the relevant bundle  is $L^q$. This generalises the appearance of $L_a\otimes L^\vee_b$ in (\ref{externcoh}) for bifundamental states.

Even though derived originally in the perturbative Type IIB context, these expressions continue to hold for general 7-branes in F-theory. This  can be derived e.g.\ with the help of the eight-dimensional twisted field theory introduced in \cite{Beasley:2008dc}.

\subsubsection*{Localised matter}

From the intersecting brane perspective it is natural to expect that additional massless matter appears at the intersection of two 7-branes.
What happens in F-theory is that as two singular loci collide transversely, the singularity type of the elliptic fibration  enhances over the intersection locus \cite{Katz:1996xe}. This intersection occurs in complex codimension one along the base. In F-theory compactifications on Calabi-Yau 4-folds, two divisors $D_a$ and $D_b$ intersect along a complex curve $C_{ab} = D_a \cap D_b$.
The appearance of extra matter can be understood  in terms of the tree of zero-size ${\mathbb P}^1$s over each singular locus. Along the intersection locus the two sets of ${\mathbb P}^1$s unite to form the (affine) Dynkin diagram of a new gauge group $G_{ab}$. Its rank is the sum of the gauge groups $G_a$ and $G_b$ along the two divisors. Note that group theoretically there may be several types of enhancements possible, each leading to different types of matter states.
By abuse of notation one calls the group $G_{ab}$ the enhanced gauge group along the intersection locus even though in actuality there exists no ${\cal N}=1$ SYM theory with a corresponding vector multiplet. However, from the reduction of the M-theory 3-form and from wrapped M2-branes along the various ${\mathbb P}^1$s in the fiber one does find as many massless states along $C_{ab}$ as is necessary to form the adjoint of $G_{ab}$. These states include the states propagating along $D_a$ and $D_b$ but also contain extra matter localised at $C_{ab}$.
Under the group theoretic decomposition 
\bea
&& G_{ab} \rightarrow G_a \times G_b  \\
&& {\rm ad}_{G_{ab}} \rightarrow ( {\rm ad}_{G_{a}},1)   \oplus (1, {\rm ad}_{G_{b}} ) \oplus \sum (R_x, U_x) 
\eea
the localised states transform in the representation $(R_x, U_x)$. These states descend from M2-branes wrapping chains of ${\mathbb P}^1$s associated both with the Dynkin diagram of $G_a$ and $G_b$, in analogy to the chains $S_{ij}$ defined in the paragraph after (\ref{Cij}). Away from the intersection locus $C_{ab}$, such chains are not of zero size and the extra matter becomes massive. 

Note that for enhancements to non-classical groups $G_{ab}$ the irreps $(R_x, U_x)$ need not correspond to two-index representations, as would be the case for perturbative Type IIB models.
The larger set of possibilities  is again due to the multiple endpoints of multi-pronged $\tiny{ \begin{pmatrix} p   \\  q    \end{pmatrix} }$ strings, which furnish a genuine strong coupling effect. While the Chan-Paton factors of a  fundamental $\tiny{ \begin{pmatrix} 1   \\  0    \end{pmatrix} }$ string with two endpoints always give rise to two-index representations, a multi-pronged string can accomodate more general charges.

An analysis of the twisted defect SYM theory along the matter curves \cite{Beasley:2008dc} confirms that this massless matter is counted by the same expressions familiar from perturbative Type IIB models \cite{Katz:2002gh},
\bea
H^i(C_{ab}, F_x \otimes K_{C_{ab}}^{\frac{1}{2}}), \quad\quad i=0,1.
\eea
Here $i=0$ and $i=1$ respectively count chiral and anti-chiral multiplets in representation $(R_x, U_x)$, and $F_x$ is schematic for suitable combinations of gauge flux restricted onto $C_{ab}$.
The chiral index follows via Hirzebruch-Riemann-Roch as 
\bea
\chi = {\rm dim} H^0(C_{ab}, F_x \otimes K_{C_{ab}}^{\frac{1}{2}}) - {\rm dim} H^1(C_{ab}, F_x \otimes K_{C_{ab}}^{\frac{1}{2}}) = \int_{C_{ab}} F_x.
\eea
We will be more specific about the concrete chirality formulae in the context of the spectral cover construction of gauge flux in section \ref{spectral_cover}.
 
The type of singularity enhancement can be read off from the discriminant locus with the help of Tate's algorithm.
As a non-trivial example we consider again the $SU(5)$ gauge theory along the divisor $S$ introduced above.
In this case the localised matter states arise at the intersection of the $SU(5)$ brane  $S$ with the $I_1$-locus $S_1$.
Since the latter carries no non-abelian gauge group, the singularity gets enhanced by rank one.
Two types of rank-one enhancements are possible for $SU(5)$ \cite{Donagi:2008ca,Beasley:2008dc}:
\begin{itemize}
\item
Enhancement $A_4 \rightarrow D_5$ corresponding to $SU(5) \rightarrow SO(10)$. The extra matter transforms in the  ${\bf 10}$ representation of $SU(5)$ as can be determined from the branching rule
\bea
{\bf 45} \rightarrow ({\bf 24})_0 + ({\bf 1})_0 + {\bf 10}_2 + {\bf \ov{10}}_{-2}. 
\eea
The subscripts denote the (formal) $U(1)$ charges under the decomposition $SO(10) \rightarrow SU(5) \times U(1)$.

From Tate's algorithm this $D_5$ enhancement occurs along the curve
\beq \label{curve10}
  P_{10}: \quad w=0 \quad \cap \quad \mathfrak{b}_5 = 0,
\eeq
along which the discriminant \eqref{Delta_SU5} scales like $w^7$, as required for a $D_5$ singularity. 
\item
Enhancement $A_4 \rightarrow A_5$ corresponding to $SU(5) \rightarrow SU(6)$. The $SU(6)$ locus hosts 
matter in the ${\bf 5}$ arising via
\bea
{\bf 35} \rightarrow ({\bf 24})_0 + ({\bf 1})_0 + {\bf 5}_1 + {\bf \ov 5}_{-1}.
\eea
This enhancement occurs whenenver
\beq \label{curve5a}
  P_5: \quad w=0 \quad \cap \quad P = \mathfrak{b}_3^2 \mathfrak{b}_4 - \mathfrak{b}_2 \mathfrak{b}_3 \mathfrak{b}_5 + \mathfrak{b}_0 \mathfrak{b}_5^2 = 0,
\eeq
consistent with the scaling $\Delta \propto w^6$ in \eqref{Delta_SU5}. 
\end{itemize}

In addition there can appear localised GUT singlets ${\bf 1}$  on matter curves away from (but possibly intersecting) the $SU(5)$ brane $S$. These localised states appear at self-intersections of the $I_1$-part $S_1$ of the discriminant, along which the gauge group enhances to $A_1$ \cite{GW-new}.

\subsubsection*{Yukawa points}

In compactifications on Calabi-Yau 4-folds, two or more matter curves can meet in points. Here the singularity type of the fiber enhances even further by the same mechanism that leads to the enhancement along matter curves. 
Consider for example  the intersection of three matter curves $C_{ab}$, $C_{bc}$, $C_{ac}$ in one point.
Group theoretically the representations of the matter states along the three curves combine to form the adjoint ${\rm ad}_{abc}$ of an enhanced gauge group $G_{abc}$.
Even though there exists no actual gauge theory associated with $G_{abc}$,  the cubic interaction term for the adjoint of this (hypothetical) enhanced gauge group leads to Yukawa interactions for the massless matter \cite{Donagi:2008ca,Beasley:2008dc}. 
This can be argued by a closer analysis of the local gauge theory description of the geometry responsible for the enhancements.
The resulting cubic interaction in the product theory $G_a \times G_b  \times G_c$  can be determined by decomposing the triple product ${\rm ad}^3_{abc}$ into gauge invariant triple products for the irreducible representations of $G_a$, $G_b$ and $G_c$.
This picture is in perfect agreement with expectations from weakly-coupled Type IIB theory, where Yukawa couplings are known to be localised at the intersection of matter curves.

To illustrate this picture in the context of our $SU(5)$ gauge theory, we note that there exist three possible rank-two enhancements of $A_4$ to $E_6$, $D_6$ and $A_7$ with the following Yukawa structure \cite{Donagi:2008ca,Beasley:2008dc}:
\begin{itemize}
  \item The ${\bf 10 \, 10 \, 5}$ Yukawa is localized at  a point of $E_6$ enhancement $\mathfrak{b}_5 = 0 = \mathfrak{b}_4$.
  \item The ${\bf 10 \, \ov 5 \, \ov 5 }$ is localized at a $D_6$ point $\mathfrak{b}_5 = 0 = \mathfrak{b}_3$. 
\end{itemize}
Indeed $\mathfrak{b}_4 = 0 = \mathfrak{b}_5$  and $\mathfrak{b}_5 = 0 = \mathfrak{b}_3$ correspond to a single and double zero of $P_5$ in agreement with the number of ${\bf \ov 5}$ representations appearing in the coupling.
\begin{itemize}
  \item At $P=0=R$ but $(\mathfrak{b}_4,\mathfrak{b}_5) \neq (0,0)$ the singularity type enhances to $A_6$. This realizes the coupling ${\bf  5 \, \ov 5 \, 1}$. The state ${\bf 1}$ 
represents a (possibly localised) GUT singlet.
\end{itemize}

We summarise the various codimension singularities of generic $SU(5)$ models and their physical interpretation in table \ref{tab:GaugeEnhanc}.
\begin{table}[t]
  \centering
    \begin{tabular}{lc|c|cc|ccccc|ll}
                 &   sing.                        & discr.         & \multicolumn{2}{|c}{gauge enh.}        & \multicolumn{5}{|c}{coeff.~vanish.~deg} & \multicolumn{2}{|c}{object}  \\
                 &   type                         & $\deg(\Delta)$ & type & group  & $a_1$ & $a_2$ & $a_3$ & $a_4$ & $a_6$${}_\big.$ & \multicolumn{2}{|c}{equation} \\ \hline\hline
      GUT:       & $\mathrm{I}_5^{\,\mathrm{s}}$  & 5 & $A_4$ & $SU(5)$    & 0 & 1 & 2 & 3 & 5 & $S:$ & $w=0$${}_\big.$${}^\big.$ \\
      matter:    & $\mathrm{I}_6^{\,\mathrm{s}}$  & 6 & $A_5$ & $SU(6)$    & 0 & 1 & 3 & 3 & 6 & $P_5:$ & $P=0$ \\
                 & $\mathrm{I}_1^{*\,\mathrm{s}}$ & 7 & $D_5$ & $SO(10)$   & 1 & 1 & 2 & 3 & 5 & $P_{10}:$ & $b_5=0$${}_\big.$ \\
      Yukawa:{\!\!\!\!\!} & $\mathrm{I}_2^{*\,\mathrm{s}}$ & 8 & $D_6$ & {\!\!\!}$SO(12)^*$ & 1 & 1 & 3   & 3 & 5 & \multicolumn{2}{|l}{$\mathfrak{b}_3=\mathfrak{b}_5=0$} \\
                 & $\mathrm{IV}^{*\,\mathrm{s}}$  & 8 & $E_6$ & $E_6$      & 1 & 2 & 2 & 3 & 5 & \multicolumn{2}{|l}{$\mathfrak{b}_4=\mathfrak{b}_5=0$}${}_\big.$ \\
      extra:     & $\mathrm{I}_7^{\,\mathrm{s}}$  & 7 & $A_6$ & $SU(7)$    & 0 & 1 & 3 & 4 & 7 & \multicolumn{2}{|l}{$P=R=0,$} \\
                 &                                &   &       &            &   &   &   &   &   & \multicolumn{2}{|l}{$(\mathfrak{b}_4,\mathfrak{b}_5)\not=(0,0)$}
    \end{tabular}
  \caption{\small Relevant gauge enhancements in a generic $SU(5)$ GUT geometry (borrowed from \cite{Blumenhagen:2009yv}).} 
  \label{tab:GaugeEnhanc}
\end{table}

\subsection{F-theory-heterotic duality}
\label{Fhet}

\subsubsection*{Generalities on the duality}

Thus far we have approached F-theory via Type IIB orientifolds and via duality with M-theory. In this section we highlight some  aspects of the duality with the heterotic string. This duality is a particularly fruitful source of inspiration for many applications to concrete model building.

The basic assertion is that F-theory on an elliptic $K3: T^2 \rightarrow {\mathbb P}^1$ is dual to the heterotic  string on $T^2$ \cite{Vafa:1996xn}. This can be argued  by comparing the moduli spaces of both theories. In particular, the heterotic string coupling ${\rm exp}(2\phi)$ is dual to the volume of the base $\mathbb P^1$ of the F-theory $K3$.

This basic duality can be extended "adiabatically" by fibering an elliptic $K3$-surface and, respectively, an elliptic curve over a common $n$-complex dimensional base $B_n$ in such a way as to produce Calabi-Yau manifolds on either side of the duality. Thus F-theory on a $K3$-fibered Calabi-Yau $(n+2)$-fold $Y_{n+2}: K3 \rightarrow B_n$ is dual to the heterotic string on the elliptic Calabi-Yau $(n+1)$-fold $Z_{n+1}: T^2 \rightarrow B_n$.
Since $Y_{n+2}$ exhibits a double fibration structure as a $K3$-fiberation over $B_n$ and, as always, as an elliptic fibration $Y_{n+2}: T^2 \rightarrow B_{n+1}$, the base $B_{n+1}$ is by itself  $\mathbb P^1$-fibered, $B_{n+1}: \mathbb P^1 \rightarrow B_n$. Therefore only a very special subclass of F-theory compactifications possesses a straightforward heterotic dual in terms of an elliptic fibration $Z_{n+1}$.

On the heterotic side, a perturbative $SO(32)$ or $E_8 \times E_8$ gauge group descends from the respective ten-dimensional theory by compactification on a smooth Calabi-Yau. If present, singularities on the heterotic compactification space lead to extra gauge group factors not related to this perturbative gauge group which can be of potentially huge rank and which are due to massless non-perturbative states. In the sequel we will focus on $E_8 \times E_8$ heterotic models with only the perturbative  group, i.e. on models for which the elliptic fibration $Z_{n+1}$ is smooth. 
Heterotic models on smooth Calabi-Yau spaces are defined in terms of a holomorphic vector bundle $V_1 \oplus V_2$ with structure group $H_1 \times H_2$ embedded into $E_8 \times E_8$. This breaks the gauge group down to the commutant $G_1 \times G_2$. The vector bundle data must therefore map into the singularity structure of the elliptic fibration $Y_{n+2}$ on the dual F-theory side.

Heterotic/F-theory duality is particularly powerful in compactifications to six dimensions, where it relates the heterotic string on a $K3$ which is elliptically fibered over ${\mathbb P}^1$ to F-theory on a $K3$-fibration over that same $\mathbb P^1$. This implies that the base $B_2$ of the F-theory elliptic fibration is itself a $\mathbb P^1$-fibration over $\mathbb P^1$. Such a fibration is called a rationally ruled or Hirzebruch surface ${\mathbb F}_k$, where $k=0,1,\ldots$ determines the structure of the fibration. The geometry of elliptic fibrations over ${\mathbb F}_k$  and applications to F-theory/heterotic duality in six dimensions have been discussed in detail in the pioneering \cite{Morrison:1996na} (more general elliptic 3-folds are analysed in \cite{Morrison:1996pp}).
The value $k$ of the F-theory base ${\mathbb F}_k$ determines the second Chern class or instanton number of the heterotic bundle $V_1 \times V_2$ embedded into $E_8 \times E_8$ as $(12-k, 12+k)$, $k = 0, 1, \ldots 12$.
The heterotic dilaton is now related to the volume of the fiber and of the base ${\mathbb P}^1$ of the F-theory $B_2$ as
\bea
{\rm exp}(2 \phi) = \frac{{\rm vol}_f}{{\rm vol}_b} .
\eea

For compactifications to four dimensions, the set of possible F-theory $K3$-fibrations is constrained by the fact that only a small number of complex base spaces $B_2$ allow for elliptic Calabi-Yau spaces $Z_{3}: T^2 \rightarrow B_2$ as required on the heterotic side. In fact, for smooth heterotic compactifications with ${\cal N}=1$ supersymmetry $Z_3$ must have $SU(3)$ holonomy and $B_2$ can only be a (blow-up in $r$ points of) $\mathbb P_2$, a (blow up of) ${\mathbb F}_k$ or the Enriques surface $K3/{\mathbb Z}_2$ \cite{Grassi-het} (see \cite{Morrison:1996pp} for more explanations).
The base $B_3$ of the dual F-theory elliptic 4-fold is then a $\mathbb P^1$-fibration over these complexes surfaces.
Such a fibration is characterised by a line bundle ${\cal T}$ over $B_2$ with first Chern class $c_1({\cal T}) = t$ as follows: Consider the rank 2 bundle ${\cal O} \oplus {\cal T}$ over $B_2$. Its fiber consists of two copies of $\mathbb C$, the fibers of the trivial line bundle ${\cal O}$ and  of ${\cal T}$. In the same manner as one forms a ${\mathbb P}^1$ by projectivising two complex coordinates $ (0,0) \neq (z_1, z_2) \simeq \lambda (z_1, z_2)$, a ${\mathbb P}^1$-fibration can be obtained as the projectivisation $ \mathbb P({\cal O} \oplus {\cal T}) =B_3$. I.e. one projectivises each fiber ${\mathbb C} \oplus {\mathbb C}$ and then fibers over $B_2$.
Let $r= c_1({\cal O}(1))$, where by abuse of notation ${\cal O}(1)$ is the line bundle over $B_3$ that reduces, along each $\mathbb P^1$ fiber, to ${\cal O}_{{\mathbb P}^1}(1)$.
Then one can show that \cite{Friedman:1997yq}
\bea
\label{rplust}
r(r+t) =0, \quad\quad c_1(B_3) = c_1(B_2) + 2r + t .
\eea
The upshot is that the class $t$ generalises the integer $k$ characterising the Hirzebruch surface ${\mathbb F}_k$ for F-theory in 6 dimensions. In oder to understand the generalisation of the relation between $k$ and the instanton number $(12-k, 12+k)$ of the heterotic bundle, we need to familiarize ourselves with the construction of vector bundles on elliptic 3-folds $Z_3$. 
Thanks to the work of Donagi \cite{donagi-1997-1} and Friedman, Morgan and Witten \cite{Friedman:1997yq} a large class of holomorphic vector bundles on elliptic fibrations are known in terms of the spectral cover construction. In the sequel we give a brief summary of the basic idea and quantities characterising a spectral cover bundle.

\subsubsection*{The heterotic spectral cover construction }
\label{SCC-het}


The construction of a rank $n$ spectral cover bundle involves two concepts: that of a spectral surface ${\cal C}^{(n)}$ and of a spectral line bundle ${\cal N}$, both describable in terms of cohomological data on $B_2$.  Under F-theory/heterotic duality this data maps to the geometry of the singular $Y_4$ and to gauge flux $G_4$.

The basic idea of the spectral cover construction is to first construct a stable $(S)U(n)$  bundle on the elliptic fibre over each point of the base and to then extend it over the whole manifold $Z_{3}$ by gluing the data together suitably. 
Recall that in general, an $(S)U(n)$  bundle defines a rank $n$ complex vector bundle. Its restriction to the elliptic fiber $E_b$ over $b \in B_2$ can be shown to be isomorphic to the direct sum of $n$ complex line bundles
\bea
{V}|_{E_b}=\mathcal{N}_1\oplus\ldots\oplus\mathcal{N}_n,
\eea
each of which has to be of zero degree. If $G=SU(n)$ as opposed to $U(n)$, ${ V}|_{E_b}$ must in addition be of trivial determinant, i.e. $\bigotimes_{i=1}^n {\cal N}_i = {\cal O}_{E_b}$. The zero degree condition on ${\cal N}_i$ implies that there exists for each  ${\cal N}_i$ a meromorphic section with precisely one zero at some $Q_i$ and a pole at $p$, the zero of the elliptic curve.  I.e.\ ${\cal N}_i = {\cal O}_{E_b}(Q_i-p)$. 
Consequently, stable $(S)U(n)$ bundles on an elliptic curve are in one-to-one correspondence with the unordered $n$-tuple of points $Q_i$, and the reduction of $U(n)$ to $SU(n)$ is encoded in the additional requirement that $\sum_i(Q_i - p) = 0$ in the group law of the elliptic curve. 

Having understood the restriction of a rank $n$ bundle ${ V}$ to each elliptic fibre, \cite{Friedman:1997yq} proceeds to construct the whole of ${V}$. In intuitive terms, the above implies that over an elliptically fibered manifold a $U(n)$ vector bundle determines a set of $n$ points, varying over the base. More precisely, the bundle ${V}$ over ${Z_3}$ with the property \cite{Friedman:1997yq}
\bea
\label{V1}
{V}|_{E_b} = \bigoplus_{i=1}^n {\cal O}(Q_i-p)
\eea
 uniquely defines an $n$-fold (ramified) cover ${\cal C}^{(n)}$ of $B_2$, the spectral cover. It is defined by a projection
\bea
\label{intpoints}
\pi_n: {\cal C}^{(n)} \rightarrow B_2 \quad\quad\quad\quad {\rm such \, \, that} \quad\quad    {\cal C}^{(n)} \cap E_b  = \bigcup_i  \,Q_i. 
\eea
That is the intersection points of ${\cal C}^{(n)}$ with the elliptic fiber are the $n$ points characterising the restriction of the bundle $V$ to the fiber. 
${\cal C}^{(n)}$ is conveniently described, as a hypersurface in ${Z_3}$, by its Poincar\'e dual two-form. The Weierstrass model $Z_{3}$ possesses a section $\sigma$ which identifies the base $B_2$ as an element of $H_4(B_2, \mathbb Z)$. This section has the important property 
\beq \label{sigma2}
  \sigma \cdot \sigma = - \sigma\,  c_1(S).
\eeq
The class of ${\cal C}^{(n)}$  can then be written as 
\bea
\label{C-class}
[{\cal C}^{(n)}] = n \sigma + \pi^*(\eta) \in H^2({Z_3}, {\mathbb Z})
\eea
for $\eta$ some effective class in $H^2(B_2, {\mathbb Z})$. In particular the first piece $n \sigma$ shows that the spectral surface is an $n$-fold cover of the base $B_2$.

Several distinct bundles over ${Z_{3}}$ may well give rise to the same spectral cover ${\cal C}^{(n)}$ since the latter only encodes the information about the restriction of ${V}$ to the fibre $E_b$. To recover ${V}$ from the spectral data we need to specify in addition how it varies over the base, i.e. ${V}|_{B_2}$. As discussed in \cite{Friedman:1997yq}  this is uniquely accomplished by the so-called spectral line bundle ${\cal N}$ on ${\cal C}^{(n)}$ with the property
\bea
\label{V2}
\pi_{n*} {\cal N} = { V}|_{B_2}.
\eea

For the first Chern class $c_1({\cal N}) \in H^2({\cal C}^{(n)}; \mathbb Z)$ of the spectral bundles one can make the general decomposition ansatz \cite{Friedman:1997yq, Andreas:2004ja}
\beq \label{c1N}
  c_1(\mathcal{N}) = \frac{r}{2} + \gamma.     
\eeq
Here we have abbreviated
\beq
  r =  {-\,c_1({{\cal C}^{(n)}})+ \pi_n^{\ast} c_1(S)} , \qquad 
  \gamma =  \frac{1}{n}\,  \pi_n^{\ast} c_1({ V})  + \gamma_u  ,
\eeq
where  $\gamma_u$ is chosen such that it satisfies $\pi_{n \ast} \gamma_u =0$. This yields 
\beq \label{gammau}
  \gamma_u=\lambda\, (n\sigma-\pi_n^{\ast}\eta +{n}\pi_n^{\ast}c_1(S)) ,
\eeq
for a number $\lambda \in \mathbb{Q}$ subject to certain constraints to be discussed shortly. Let us further parametrize $c_1({ V})$ by some element  $\zeta \in H^2(S; {\mathbb Z})$ \cite{Andreas:2004ja},
\beq
  \zeta = c_1({V}) .
\eeq
Note that this bundle exists for generic complex structure since it only involves  $\sigma$ and the pullback of classes from $B_2$.
The parameter $\lambda \in \mathbb Q$ has to be chosen such that $c_1({\cal N})$ defines an integer class in $H^2({\cal C}^{(n)}; \mathbb Z)$. On the non-Calabi-Yau space $X$ the adjunction formula leads to
\beq \label{c1Cn}
  - c_1({{\cal C}^{(n)}}) = (n-2) \sigma + \pi_n^* ( \eta -2 c_1(B_2)).
\eeq
Putting everything together, we have 
\bea \label{linebundle}
    c_1(\mathcal{N}) = {} & {} - \sigma + n\left({\textstyle  \frac{1}{2}+\lambda }\right)\,\sigma + \left({\textstyle \frac{1}{2} - \lambda}\right) \pi_n^\ast \eta \\
    & {} + \left( {\textstyle - \frac{1}{2}+ n \lambda} \right) \pi_n^\ast c_1(B_2) + {\textstyle \frac{1}{n}}\, \pi_n^\ast \zeta.
 \eea

E.g. for an $SU(5)$ bundle, integrality of $c_1({\cal N})$ puts the value of $\lambda \in {\mathbb Q}$ subject to the constraints 

\beq \label{Ninteger}
    5\left ({\textstyle \frac{1}{2}+  \,\lambda }\right)  \in {\mathbb Z}\ ,\qquad 
    \left({\textstyle \frac{1}{2}-\lambda}\right)\, \eta +\left({\textstyle 5 \lambda - \frac{1}{2}}\right) c_1(S) \in H^2(S; {\mathbb Z})\ .
\eeq
From these data one can compute the higher Chern classes. For an $SU(n)$  bundle these are  \cite{Friedman:1997yq,Curio:1998vu}
\bea
   \int_{B_2} c_2(V) &=& \int_{B_2} \eta \sigma - \frac{1}{24}  \chi_{SU(n)} - \frac{1}{2} \int_{B_2} \pi_{n*}(\gamma^2), \\
    \int_{Z_3} c_3(V)  &=&  \lambda \, \eta (\eta - n c_1(B_2)), \nonumber
\eea
where  $\chi_{SU(n)}  =\int_{B_2} c_1^2(B_2) (n^3-n) +3n\, \eta \big(\eta-n c_1(B_2)\big)$. \\ \vspace{1mm}

To summarize, a $U(n)$ spectral cover bundle is characterised by the following topological data:
\begin{itemize}
\item
the class of the spectral surface $[{\cal C}^{(n)}] = n\, \sigma + \pi^* \eta$,  $\eta \in H^2(B_2, \mathbb Z)$;
\item
the first Chern class of the spectral line bundle $c_1({\cal N})$ as in  (\ref{c1N}).
\end{itemize}

Under F-theory-heterotic duality the bundle data map partly into the singular geometry of $Y_4$ and partly into gauge flux $G_4$.
The gauge groups $G_1$ and $G_2$ are localised on the divisors given by the base $B_2$ located at the north and south pole of the $\mathbb P^1$ that constitutes the fiber of $B_3: \mathbb P^1 \rightarrow B_2$.
\begin{itemize}
\item
The classes $\eta_i$ for the two bundles embedded into $E_8^{(1)}$ and $E_8^{(2)}$ correspond on the F-theory side to \cite{Friedman:1997yq} 
\beq \label{eta12}
  \eta_1 = 6 c_1(B_2) - t\ , \qquad \eta_2 = 6 c_1(B_2) + t,
\eeq
where $t = c_1({\cal T})$, see the discussion around (\ref{rplust}).
\item
The quantity $\gamma_u$ appearing in (\ref{gammau}) maps into $G_4$ flux; in particular \cite{Curio:1998bva}
\bea
\int_{\ov Y_4} G_4 \wedge G_4 = - \int_{B_2} \pi_{n*}(\gamma_1^2 + \gamma_2^2) .
\eea
\end{itemize}
In \cite{Friedman:1997yq}, also methods for the construction of gauge bundles with more general structure group including $E_6, E_7, E_8$ are developed. If one embeds a bundle with structure group $H_1 \times E_8$ into $E_8 \times E_8$,  the visible heterotic gauge group is $G=G_1$.  On the F-theory side this maps into a single gauge group $G_1$ along the base of the $\mathbb P^1$-fibration $B_3$. Note that this is precisely the structure encountered for global Tate models.


\subsection{The spectral cover construction for F-theory models}\label{spectral_cover}

Let us come back to general F-theory models on an elliptic fibration $Y_4$.
As we have reviewed, the global structure of the Weierstrass model $Y_4$ is specified by the sections $f \in H^0(B_3, K_{B_3}^{-4} )$ and $g \in H^0(B_3, K_{B_3}^{-6})$. Locally this can  be brought into the form (\ref{Tate1}) of a Tate 
model. A special class of fibrations is even globally of the Tate form and thus based on an underlying $E_8$ gauge symmetry, broken to gauge group $G$ along a single divisor $S$.

If one is interested not in the full details of the global 4-fold geometry, but merely in the physics on the divisor $S$, one can restrict the Tate model to the neighbourhood of $S \subset B_3$. This restriction of the Tate constraint to the neighbourhood of $S$ likewise goes under the name  spectral cover construction, whose application we just described in the context of heterotic model building. 
By heterotic/F-theory duality it is clear that spectral covers have a natural appearance also for F-theory compactifications with a heterotic dual. 
More recently, however, it has been appreciated \cite{Hayashi:2009ge, Donagi:2009ra} that  spectral covers are useful to describe the geometry and gauge flux of F-theory compactifications even without (simple) heterotic duals - at least in and to some amount even beyond a local picture.

\subsubsection*{The general philosophy}

Before discussing the technicalities, let us try and gain an intuitive understanding of the appearance of the spectral cover construction.
For general F-theory models, the  essence of the spectral cover idea is to zoom into the local neighbourhood of the divisor $S: w=0$ within $B_3$ by discarding all terms of higher power in the normal coordinate $w$ that appear in the sections  $\mathfrak{b}_n$, defined in (\ref{TateSU5b1}) for the case of $SU(5)$.
The restrictions of  $\mathfrak{b}_n$ to the divisor $S$,
\bea
\label{bi}
b_n = \mathfrak{b}_n|_{\omega=0},
\eea
are therefore sections entirely on $S$.
In this local picture  the brane $S$ is described as the base of the bundle $K_S \rightarrow S$, given by $s=0$.
 The neighbourhood of $S$ is then modelled by a spectral surface viewed as a divisor of the total space of $K_S$. In the sequel we will concentrate on the Tate model for an $SU(5)$ GUT symmetry along $S$ with associated spectral surface
 \beq \label{C5b}
\mathcal{C}^{(5)}:  \,  b_0 s^5 + b_2 s^3 + b_3 s^2 + b_4 s + b_5 = 0.
\eeq
One can think of $\mathcal{C}^{(5)}$  as encoding the information about the  discriminant locus in 
the local vicinity of $S$. 
In particular, as we will see the intersections of $\mathcal{C}^{(5)}$ and $S$ determine the ${\bf 10}$-matter curves (\ref{curve10}) on $S$. 
It is also clear from the relation (\ref{bi}), though, that all the information in $\mathfrak b_n$ 
contained in  terms higher in $w$ is lost in the spectral cover approach.

The spectral cover approach to F-theory model building serves in particular as an auxiliary construction to construct the gauge flux required for chirality of the model. Given the local nature of the spectral covers, it is reasonable to suspect that this yields a correct description of gauge flux near the brane $S$ and in particular along the matter curves on $S$. This is good news as it is the restriction of the fluxes onto these curves which governs the chirality of a model, but more work is needed to fully understand the continuation of the fluxes in a global construction.

Recall from section \ref{Fhet} that in models with a heterotic dual, the elliptic 4-fold $Y_4$ also has the structure of a $K3$-fibration ${K3} \rightarrow B_2$ over a complex surface $B_2$; i.e.\ the base space $B_3$ of the elliptic fibration $Y$ is itself  globally $\mathbb P^1$-fibered over $B_2$. If in addition on the heterotic side the second $E_8$ factor is broken by an $E_8$ bundle, then the only non-abelian gauge group is localised along $B_2$, which in the $SU(5)$ example we would identify with the GUT divisor $S$. The GUT divisor is therefore the base of a \emph{globally} defined fibration in models with heterotic dual. 
In general F-theory models, we have seen that this is not the case. However, from the discussion of the split resolution in section \ref{sec_Gau} we know that one can \emph{locally} view $S$ as the basis of an ALE fibration which describes the singularity structure along $S$. The ALE fiber contains a distinguished set of two-cycles $\Gamma_{E_8}^i$ whose intersection form is related to the Cartan matrix of $E_8$. For generic non-zero size of these two-cycles the $E_8$ symmetry is broken. If the divisor $S$ exhibits enhanced gauge symmetry $G$ this is because some of the $\Gamma_{E_8}^i$, called $\Gamma_G^i$ in the sequel, in the fiber shrink to zero size. The intersection matrix of the two-cycles $\Gamma_H^i$ with non-zero size is related to the Cartan matrix of the commutant $H \subset E_8$ of $G$.
 
This picture is of course very reminiscent of the breaking of $E_8$ to  $G$ in the heterotic string by means of a gauge bundle with structure group $H$. Furthermore, as seen at the end of section \ref{Fhet}, in F-theory some of the degrees of freedom of the heterotic vector bundle are encoded purely geometrically, while others map to gauge flux. The geometric part is interpreted in the local field theory of \cite{Beasley:2008dc} as encoding the vacuum expectation value of the  Higgs field $\varphi \in H^{0}(S, K_S)$ associated with the normal fluctuations of the 7-brane. The spectral cover now is designed to describe the size of the non-zero two-cycles responsible for the breaking of $E_8$ to $G$ along $S$.

Finally we stress once again that for the special case of global Tate models the underlying $E_8$ structure is exact. It is clear then that the spectral cover construction, which geometrises the breaking of $E_8$ to $G$ via the Higgs bundle of structure group $H$, has a chance to capture more than just the very local neighbourhood of the divisor $S$.

\subsubsection*{Technical details of spectral covers for an \boldmath$SU(5)$ model} 
In what follows we restrict ourselves to the spectral cover description of a $G=SU(5)_{GUT}$ singularity along a divisor $S \subset B_3$.  The complement of $SU(5)_{GUT}$ in $E_8$ is denoted by $H= SU(5)_{\perp}.$ For  more  background and for details on more general configurations we refer to \cite{Hayashi:2009ge, Donagi:2009ra}. 

The starting point is to construct an auxiliary non-Calabi-Yau 3-fold ${\cal W}$ as a fibration over $S$ which encodes the singular geometry of $S$ in $B_3$. This space ${\cal W}$ serves as a compactification of the total space of the bundle $K_S \rightarrow S$ introduced before. We will therefore think of $S$ either as a divisor on $B_3$ or as the base of a fictitious 3-fold ${\cal W}$. The definition of ${\cal W}$ is as the projectivized bundle over the GUT divisor $S$
\beq \label{defX}
  {\cal W} = {\mathbb P} ({\cal O}_{S} \oplus K_S) ,\qquad p_{\cal W}: {\cal W} \rightarrow S,
\eeq
where $p_{\cal W}{\cal W}$ is the projection to the base of the bundle. The base $S$ is viewed as the vanishing locus of the section $\sigma$ in ${\cal W}$. This section satisfies the important relation
\beq \label{sigma2}
  \sigma \cdot \sigma = - \sigma\,  c_1(S).
\eeq
This should ring a bell and remind us of the relation (\ref{sigma2}) encountered for the heterotic spectral cover construction on the physical elliptic Calabi-Yau space $Z_3$.
Indeed from now on the construction formally proceeds in the same fashion as described in section \ref{SCC-het}, replacing the base $B_2$ of the elliptic Calabi-Yau fibration $Z_3$ by the GUT divisor $S$, viewed as the base of the  ${\mathbb P}^1$-fibration ${\cal W}$. Unlike $Z_3$, ${\cal W}$ is not Calabi-Yau and has first Chern class
\beq
  c_1({\cal W}) = 2 \sigma + 2 c_1(S).
\eeq
The spectral cover is constructed as a 5-fold cover of $S$ within ${\cal W}$,
\beq
  \pi_5: {\cal C}^{(5)} \rightarrow S.
\eeq
While in the heterotic context the intersection points of ${\cal C}^{(5)}$ with the elliptic fiber encode the information about the vector bundle $V$ restricted to the fiber, see eqn.\ (\ref{intpoints}), here the $5$ intersection points with the $\mathbb P^1$-fiber denote the 5 eigenvalues of the $SU(5)_H$-valued Higgs field $\varphi$ along $S$  \cite{Hayashi:2009ge, Donagi:2009ra}, which in turn specifies the local neighbourhood of $S$ inside $B_3$ \cite{Beasley:2008dc,Donagi:2009ra}.

Given the implicit underlying $E_8$ structure of the ALE fibration, the massless matter representations of $G= SU(5)$ can be understood as the irreducible representations $R_x$ in the decomposition ${\bf 248} \rightarrow \sum {\rm ad}_i + \sum_x (R_x, U_x)$, 
\bea \label{matter-decomp-SU(5)}
  {\bf 248} \mapsto ({\bf 24},1) + (1,{\bf 24}) + [ ({\bf 10}, {\bf 5}) + ({\bf \ov 5}, {\bf 10}) + h.c.].
\eea
The matter curve $P_{10}$ is the locus $b_5=0$ on $S$. It is associated with the spectral cover in the fundamental representation of $H= SU(5)_{\perp}$ because the ${\bf 10}$ appears as $ ({\bf 10}, {\bf 5})$ in (\ref{matter-decomp-SU(5)}). Let us also define the object ${\cal P}_{10}$ viewed as   a curve in ${\cal W}$ \cite{Donagi:2009ra},
\beq
  {\cal P}_{10} =  {\cal C}^{(5)} \cap \sigma \subset {\cal W}.
\eeq
Then the matter curve on $S$ is related to ${\cal P}_{10}$ as
\beq
  [P_{10}] = [{\cal P}_{10} ] |_{\sigma} =  (5 \sigma +\pi_5^*  \eta)|_{\sigma} = \eta - 5 c_1(S)
\eeq
with the help of \eqref{sigma2}, i.e.\ the restriction of ${\cal P}_{10}$ is cohomologically equivalent to the matter curve $P_{10}$. 

The matter curve for the ${\bf \ov  5}$ on $X$ is more complicated and was analyzed in detail in the context of the heterotic string in \cite{Donagi:2004ia, Blumenhagen:2006wj, Hayashi:2008ba}, to which we refer for more details.

\subsubsection*{Gauge flux in  $SU(5)$ GUT models}

So far all that the spectral cover approach has done for us is to rewrite the geometric data in a seemingly more complicated manner. Its actual power, however, becomes apparent once one includes gauge flux into the compactification. 
In the context of $SU(5)$ GUT theories one distinguishes between the gauge flux along the GUT divisor $S$ itself and the flux along the matter branes which constitute the  ${I}_1$ locus of the discriminant. The first type of gauge flux takes values in $G=SU(5)_{GUT}$  and will thus further break the $SU(5)$ GUT symmetry. As for flux on the matter branes, the spectral cover approach makes use of the fact that ${\cal C}^{(5)}$ describes the geometry in the vicinity of $S$ and in particular the \emph{local} geometry of the ${I}_1$ component.  
In the machinery of the spectral cover, gauge flux on ${I_1}$  is therefore given in terms of the so-called spectral line bundle ${\cal N}$ along ${\cal C}^{(5)}$ defined by its first Chern class
\beq
  c_1({\cal N}) \in H^{(1,1)}({\cal C}^{(5)}; \mathbb Z) .
\eeq  
Since ${\cal C}^{(5)}$ is a 5-fold cover of $S$ in ${\cal W}$ one can push this line bundle forward to $S$ via ${\pi_5}_\ast$. This defines a rank-5 vector bundle 
\beq
  V = \pi_{5 \ast} {\cal N} 
\eeq
on $S$. Following the general logic of the ALE fibration over $S$ the structure group of this bundle $V$ is identified with the commutant $SU(5)_{\perp} \subset E_8$ of the GUT $SU(5)$ along $S$ and therefore indeed associated with the flux on $I_1$.

Therefore the construction of gauge flux on the matter branes is reduced to the problem of constructing an $SU(5)_H$ bundle $V$ on the basis of the ${\mathbb P}^1$-fibration ${\cal W}$.
At this point we can again refer to the discussion of the spectral line bundle and its associated spectral cover bundle in the heterotic context, which proceeds, \emph{mutatis mutandis}, in the same manner.
The spectral cover approach thus provides a concrete dictionary to describe a certain subclass of gauge fluxes in terms of a few parameters. Moreover, it allows for the computation of the chiral index for the various matter representations localised along the matter curves in terms of these input parameters and is therefore of considerable practical use for model building purposes.
It would lead too far to derive these formulae for the chirality in full detail here. We simply state without proof that the chiral index of states in the ${\bf 10}$ representation localised on the curve $P_{10}$ on $S$ is given by \cite{Donagi:2009ra},
\beq \label{chi10a}
  \chi_{\bf 10} =\sigma \cdot {\cal C}^{(5)} \cdot \gamma_u =  - \lambda  \int_S  \eta  \underbrace{( \eta - 5 c_1(S))}_{P_{10}},
\eeq
with $\gamma_u$ as in eqn.\ (\ref{gammau}) for $SU(5)_H$ bundles.
Consistently, this is also the number of chiral families in the  ${\bf \ov 5}$ representation.

Let us conclude this general presentation of the spectral cover construction by taking up our initial concerns about its global validity.
We have stressed several times that as of this writing it yet remains to find a general description of gauge flux in terms of global $G_4$ flux. Nonetheless it is remarkable that the spectral cover construction does capture some of the global aspects of the geometry correctly. More precisely this is the case for the special subclass of elliptic fibrations that can globally be described as a Tate model with non-abelian gauge group $G \subset E_8$ solely along a divisor $S$. If we take the spectral cover construction at face value for a second one can conjecture a simple closed expression for the Euler characteristic of the resolution $\ov Y_G$ of the singular 4-fold given by \cite{Blumenhagen:2009yv}
\beq \label{chi-prop_app}
  \chi(\ov Y_G) = \chi^*(Y) + \chi_{H} - \chi_{E_8}. 
\eeq
Here $\chi^*(Y)=  12 \int_B c_1(B_3)\, c_2(B_3) + 360 \int_B c^3_1(B_3)$ denotes the expression valid for a smooth elliptic fibration over the base $B_3$, and the remaining two terms are listed in table \ref{chigaugegroups}. They involve data solely on $S$ and reflect the underlying $E_8$ structure of the spectral cover construction, in which we think of first enhancing the singularity over $S$ to $E_8$ and then breaking it via the $H$-bundle down to $G$. 
This formula can be derived for models with a heterotic dual by F-theory-heterotic duality. 
The point is now that for global Tate models the underlying $E_8$ structure is correct globally. Indeed for global Tate models in which an explicit resolution of the singularities is available, the result of (\ref{chi-prop_app}) can be compared to the value of $\chi(\ov Y_G)$ computed explicitly via resolution. This has been performed in \cite{Blumenhagen:2009yv,Krause} for a number of cases using the machinery of toric geometry, finding perfect agreement.\footnote{To be precise, the story is more complicated: The described match between the local spectral cover and the global Euler characteristic has been established for {\emph{generic}} Tate models corresponding to the \emph{non-split} spectral cover construction outlined above. As will be discussed in section \ref{proton} for many phenomenological applications, $U(1)$ selection rules are desirable. In \emph{local}  language this leads to the concept of a \emph{split} spectral cover \cite{Marsano:2009gv}, and the simple generalisation of (\ref{chi-prop_app}) to these cases, applied in the models of \cite{Marsano:2009ym,Blumenhagen:2009yv,Marsano:2009gv,Krause,Marsano:2009wr,Chen:2010ts}, has not been proven to match the result of a direct \emph{global} computation because such split spectral cover models cannot directly be implemented as Tate models. In \cite{GW-new}, so-called $U(1)$ restricted Tate models were introduced as the correct \emph{global} framework to implement $U(1)$ symmetries. For these a direct computation of the Euler characteristic is possible e.g. via toric methods.}

For such geometries it is not unreasonable that also the gauge flux constructed via the spectral covers has a global extension. It is this expectation that underlies the global F-theory models as existent in the literature as of this writing, but a more complete understanding of gauge flux without relying on the spectral cover construction is desirable. For example \cite{Marsano:2010ix} proposes a global extension of spectral cover fluxes in terms of a so-called spectral divisor. This is designed in such a way as to reproduce the expressions for the chirality of states along the GUT branes, i.e. the local aspects associated with non-abelian gauge symmetry. It remains to be seen how to capture also genuinely global aspects of the flux such as the integral $\int_{\ov Y_4} G_4 \wedge G_4$  or the chiral index of singlets under the non-abelian group $G$.

\begin{table}[ht] 
  \renewcommand{\arraystretch}{1.3} 
  \centering
  \begin{tabular}{c|c|c} 
    $G = E_8/H$ & $H$ & $\chi_H$$_\big.$      \\ 
    \hline\hline 
    $E_{9-n}, \ n\le 5$ & $SU(n)$ &  $\int_S c_1^2(S) (n^3-n) +3n\, \eta \big(\eta-n c_1(S)\big)$$^\big.$ \\
    $SU(3)$    & $E_6$ & $72 \int_S  \bigl( \eta^2 - 7\eta c_1(S) + 13 c_1^2(S)\bigr)$ \\
    $SU(2)$  &  $E_7$ & $18 \int_S \bigl( 8\eta^2 - 64\eta c_1(S) + 133 c_1^2(S)\bigr)$ \\
    - & $E_8$ & $120 \int_S  \bigl( 3\eta^2 - 27\eta c_1(S) + 62 c_1^2(S)\bigr)$ \\
  \end{tabular} 
  \caption{\small Redefined Euler characteristic for $E_n$-type gauge groups. Here $\eta$ is given by $\eta =6c_1(S) + c_1(N_S)$.}
  \label{chigaugegroups}
\end{table}

\section{Phenomenological applications to GUT model building}
\label{sec_Phe}

As stressed in the introduction, F-theory combines two characteristic aspects of Type II orientifolds with D-branes on the one hand and of heterotic string vacua on the other which are of general phenomenological interest and of significant use in the context of realistic GUT model building. These are
\begin{itemize}
\item the appearance of exceptional gauge groups as in $E_8 \times E_8$ heterotic string constructions and
\item the localisation of gauge degrees of freedom, matter states and Yukawa interactions as in perturbative D-brane models.
\end{itemize}

Beginning with \cite{Donagi:2008ca,Beasley:2008dc,Beasley:2008kw,Donagi:2008kj}  the prospects of F-theory compactifications for GUT phenomenology have recently been under intense investigation.
While in principle applicability of F-theory  is by no means restricted to GUT models, the relevance of exceptional gauge groups in the context of unification singles out this class of constructions as the one with the most distinctive F-theoretic features as compared with intersecting brane models in perturbative orientifolds.  Promising starting points for the construction of GUT models are the gauge groups $SO(10)$ and $SU(5)$. Most efforts in the F-theory literature have focused on the latter, mainly due to complications with GUT breaking via internal fluxes.

Given the localisation of gauge degrees of freedom in F-theory, a considerable number of phenomenological questions can be discussed already at the level of local models.
Among these are the structure of the GUT matter curves  and the details of GUT matter Yukawa couplings.
Several other issues, by contrast, can only be addressed in a satisfactory manner in the context of a globally defined compactification. This is true in particular for all aspects of $U(1)$ symmetries - including $SU(5)$ GUT breaking via hypercharge flux and abelian selection rules - and the physics of GUT singlets (e.g.\ certain aspects of neutrino physics), which are localised away from the GUT divisor.

It is beyond the scope of these lectures to survey all exciting aspects of F-theory GUT model building that have emerged recently; rather we  will outline some of the general philosophy behind the construction of $SU(5)$ GUT models. Armed with this background the interested reader can easily delve into more advanced topics.

\subsection{SU(5) GUT models and the principle of decoupling}
\label{sec_SU(5)}

$SU(5)$  is the mother of all GUT groups. In Georgi-Glashow $SU(5)$ models  \cite{Georgi:1974sy}, the embedding of the MSSM gauge group $SU(3) \times SU(2) \times U(1)_Y$ rests on the identification of the $U(1)_Y$ generator with the Cartan generator $T={\rm diag}(2,2,2,-3,-3)$ within $SU(5)$. The MSSM matter is organised into $SU(5)$ multiplets 
as
\bea
\label{GGmatter}
&& {\bf 10} \leftrightarrow (Q_L,u_R^c, e_R^c), \quad\quad {\bf \ov 5_m} \leftrightarrow (d_R^c, L), \quad\quad {\bf 1} \leftrightarrow \nu_R^c, \nonumber \\
&& {\bf 5_H} \leftrightarrow (T_u, H_u) ,\quad\quad  {\bf \ov 5_H} \leftrightarrow (T_d, H_d). 
\eea
The triplets $T_u, T_d$, which are not present in the MSSM, must receive high-scale masses via doublet-triplet splitting.
An alternative $SU(5)$ GUT scenario called flipped $SU(5)$ \cite{Barr:1981qv,Antoniadis:1987dx} starts from gauge group $SU(5) \times U(1)_X$ and the MSSM matter is related to the identifications (\ref{GGmatter})
 by "flipping" $e_R^c \leftrightarrow \nu_R^c$ and $d_R^c \leftrightarrow u_R^c$. Since $U(1)_Y$ arises as a combination of $U(1)_X$ and the Cartan generator $T={\rm diag}(2,2,2,-3,-3)$ of $SU(5)$, flipped $SU(5)$ is a unified model only if $SU(5) \times U(1)_X$ is itself embedded into a higher group such as $SO(10)$.
For definiteness we focus for now on Georgi-Glashow $SU(5)$ models.

The geometric origin \cite{Donagi:2008ca,Beasley:2008dc,Beasley:2008kw,Donagi:2008kj} of the $SU(5)$ GUT gauge group, the matter representations and the Yukawa couplings in terms of codimension one, two and three singularities of the elliptic fibration has been discussed in sections \ref{sec_Tat} and \ref{matter-gen}, see table \ref{tab:GaugeEnhanc} for a summary.  Matter charged under $SU(5)$ localises on the curves $P_{\bf 10}$ and $P_{\bf 5}$ on the GUT divisor $S$, while the role of $\nu_R^c$ can be played by any GUT singlets participating in the coupling ${\bf 5_H \, \ov 5_m \, 1}$. Note that for generic $SU(5)$ geometries, the matter curve for the ${\bf 5}$ representation is a single connected object; in this situation all three generations of $\bf \ov 5_m$ and the vector-like pair $\bf{ 5_H + \ov 5_H}$ are localised on the same curve. We will see later that this is unacceptable for phenomenological reasons and the geometry must be further refined.

The MSSM Yukawa couplings follow from the $SU(5)$ GUT interactions by decomposition of the $SU(5)$ representations as 
\bea
{\bf 10 \, 10 \, 5_H} \longrightarrow Q_L \, u_R^c \, H_u, \quad {\bf 10 \,  \ov 5_m \, \ov 5_H} \longrightarrow L \, e_R^c H_d +  Q_L \, d_R^c \, H_d .
\eea
The  ${\bf 10 \, 10 \, 5_H}$ Yukawa coupling localises at a point of $E_6$ singularity, which is a strong coupling phenomenon. It is for the sake of this coupling that exceptional symmetry is essential. In perturbative Type II orientifolds this interaction is forbidden by global $U(1)$ selection rules and can only be generated by D-brane instantons \cite{Blumenhagen:2007zk}. The natural presence of this crucial coupling, from which the up-quark masses descend, is the prime motivation to pursue F-theory as the framework for GUT models with 7-branes; in perturbative models, the volume of the D-brane instanton generating  the ${\bf 10 \, 10 \, 5_H}$ Yukawa coupling would have to be tuned so as to prevent too large suppression, see \cite{Blumenhagen:2008zz} for recent examples. On the other hand, in flipped $SU(5)$ models, where the ${\bf 10 \, 10 \, 5_H}$ coupling  gives rise to down quark masses, such a non-perturbative suppression can be a welcome rationale to argue for the hierarchy between the top and bottom quark mass.

An important question concerns the nature of the GUT brane $S$, which must be a K\"ahler surface embedded into the base space $B$ as a holomorphic divisor. In full generality no water-proof restrictions on $S$ can be given other than it had better support one of the GUT breaking mechanisms which will be discussed in section \ref{GUTbreaking}.
A  reasonable, though not strictly necessary organising principle, however, is to require the existence of a well-defined decoupling limit for gravity \cite{Beasley:2008kw}. This paradigm is inspired by the separation of the four-dimensional GUT scale $M_{\rm GUT} = 10^{16} \,  {\rm GeV}$ and the Planck scale $M_{Pl.} = 10^{19}\,  {\rm GeV}$ together with UV completeness of GUT models.

The four-dimensional Planck scale arises from dimensional reduction of the Einstein-Hilbert term in Einstein frame,
\bea
S_{\rm EH}= M_\ast^8 \int_{\mathbb R^{1,3} \times B} \sqrt{-g} R \quad   \Rightarrow \quad M^2_{Pl.} = M^8_{\ast} \, {\rm Vol}(B). 
\eea
Here $M_{\ast}$ is the M-theory fundamental length scale inherited via M/F-theory duality and can be viewed, in the IIB limit, as the value of $\ell_s^{-1}$ in the Einstein frame.
On the other hand, the GUT scale, determined by the breaking scale of $SU(5)$ down to the Standard Model gauge group, is parametrically given by the volume of the GUT divisor $S$, 
\bea
M^4_{\rm GUT} \simeq {\rm Vol}^{-1}(S).
\eea
 For example in the context of GUT breaking via hypercharge flux as discussed in section  \ref{GUTbreaking} this approximate relation arises because the volume of $S$ sets the flux induced mass  of the $X-Y$ gauge bosons within $SU(5)$; after all these states propagate along the whole divisor $S$. 
The observed hierarchy of about $10^{-3}$ between $M_{\rm GUT}$ and $M_{Pl.}$ translates into a small hierarchy between the typical radii of the GUT brane and the six-dimensional F-theory base of approximately
\bea
\label{estimate}
\underbrace{\ell_s}_{0.2 x }  < \underbrace{R_{S} }_{2.2 x}< \underbrace{R_{B}}_{5.6 x}, \quad\quad x= 10^{-16} \,  {\rm GeV}^{-1}.
\eea
The relation to $\ell_s$ has been chosen in such a way that in addition  
\bea
\alpha_{\rm GUT}^{-1} = M_{\ast}^4 \, {\rm Vol}(S) \simeq 24.
\eea 
This equality in turn follows parametrically by reduction of the eight-dimensional Yang-Mills action  $S_{YM} = M_{\ast}^4  \int_{{\mathbb R}^{1,3}  \times S}  F^2$.

Note that for phenomenological viability of a  model it is sufficient to stabilise the moduli in agreement with the crude estimate of (\ref{estimate}). \emph{A fortiori}, it is often postulated\cite{Beasley:2008kw,Donagi:2008kj,Cordova:2009fg} that the GUT brane $S$ allow for a limit 
\bea
{\rm Vol}(B) \rightarrow \infty, \quad\quad {\rm Vol}(S) \quad {\rm finite}
\eea
such as to decouple gravity completely - at least in principle. Alternatively to this physical decoupling limit one can consider the mathematical decoupling limit \cite{Cordova:2009fg} 
\bea
\label{mathlimit}
{\rm Vol}(S) \rightarrow 0, \quad\quad  {\rm Vol}(B) \quad  {\rm finite},
\eea
even though the two are not completely equivalent \cite{Krause}. 
This mathematical decoupling limit can be taken  if the surface $S$ is Fano, which amounts to requiring that $\int_C K_S^{-1} >  0$ for every holomorphic curve $C$ in $S$.
The list of such Fano surfaces is very restrictive and consists of the del Pezzo surfaces $\mathbb P^1 \times \mathbb P^1$, $\mathbb P^2$ and  $dP_r, r=1, \ldots,8$. The latter are defined as $\mathbb P^2$ with $r$ points in generic position blown up to $\mathbb P^1$. Since these surfaces are used extensively in the F-theory GUT literature, we briefly collect some of their basic topological properties: The second homology $H_2(dP_r,\mathbb Z)$ is spanned by the elements $l, E_1, \ldots, E_r$, where $E_i$ denote the i-th blow-up $\mathbb P^1$ and $l$ the hyperplane class inherited from $\mathbb P^2$. The non-vanishing intersection numbers are  $l\cdot l =1, E_i \cdot E_j = - \delta_{ij}$. The first Chern class of the anti-canonical bundle is given by $c_1(K_{dP_r}^{-1}) = c_1({dP_r}) = 3l - \sum_{i=1}^r E_i$ with $c_1^2(dP_r) = 9-r$ and $c_2(dP_r) = 3+r$.

 Generic Fano surfaces are shrinkable in the sense of (\ref{mathlimit}). At closer inspection, however, the existence of suitable matter curves and possible global embeddings in $SU(5)$ models poses certain restrictions on the shrinkability of the GUT divisor  and forbids shrinkability to a point in generic situations. Rather the GUT brane can only shrink to a curve or to a point in such a way that another divisor shrinks simultaneously. For more information on these restrictions we refer to  \cite{Cordova:2009fg,Krause}.

\subsection{Options for GUT breaking}
\label{GUTbreaking}

Let us now discuss possible ways to break $SU(5)$ to the observed $SU(3) \times SU(2) \times U(1)_Y$ MSSM gauge group. 
As in all brane constructions there exist three options to accomplish this:
\begin{itemize}
\item
{\bf Via a GUT Higgs} (scalar field) in the adjoint representation $\bf 24$ of $SU(5)$. This effectively realises the GUT breaking mechanism of conventional field-theoretic GUT models.
The string theoretic origin of the Higgs would be a brane deformation modulus counted by $H^{0}(S,K_S)$. An obvious challenge associated with this approach is the generation of a suitable potential for the GUT Higgs field which leads to dynamical symmetry breaking. In principle 7-brane deformation moduli are stabilised by background fluxes, but no concrete setup has been described so far that incorporates \emph{dynamical} GUT breaking.

\item
{\bf Via Wilson line moduli}. These correspond to elements of $H^{1}(S)$ and likewise transform in the adjoint of $SU(5)$. The same remarks concerning the generation of a potential for dynamical symmetry breaking apply with the added complication that Wilson line moduli are not fixed by background fluxes.
Alternatively one can consider GUT breaking by discrete Wilson lines.  These are available whenever the brane $S$ has a discrete, but non-trivial first homotopy group $\pi_1(S)$.
The VEV of the discrete Wilson line is now topological and part of the defining data of the compactification. Note that this is exactly the same strategy as pursued in Calabi-Yau compactifications of the $E_8 \times E_8$ heterotic string with $SU(N)$ gauge bundles.
In some sense this is the cleanest approach to GUT symmetry breaking. The implementation of F-theory models along GUT divisors with non-trivial $\pi_1(S)$ has not been achieved in the literature so far, but seems a promising avenue for future research.

\item

{\bf Via hypercharge flux}. This corresponds to turning on non-trivial gauge flux $F_Y$ associated with the hypercharge generator $T_Y = {\rm diag}(2,2,2,-3,-3)$.
As for Wilson lines the flux is part of the defining topological data of the compactification and circumvents the quest for the dynamical generation of a symmetry breaking scalar potential. This approach does not rest on any strong coupling effect  and is equally possible for perturbative Type IIB models. 

\end{itemize}

Note that the last two GUT breaking mechanisms are not available in conventional, four-dimensional field theoretic GUTs. Studying their possible consequences for the more detailed phenomenology of $SU(5)$ GUTs is  therefore particularly interesting. 

\subsection{Some constraints from hypercharge flux}
\label{sec_Som}

For definiteness we focus on the last GUT breaking mechanism. This is also the only possibility for del Pezzo surfaces, which have no geometric deformations and no Wilson line moduli. The use of hypercharge flux was suggested in the context of F-theory model building in \cite{Beasley:2008kw,Donagi:2008kj} and first realised in compact models with 7-branes within perturbative Type IIB orientifolds in \cite{Blumenhagen:2008zz}.
In IIB language one simply embeds a non-trivial line bundle ${ L}_Y$ along the GUT brane into $SU(5$) by identifying its structure group with $U(1)_Y$. In F/M-theory language this corresponds to $G_4$ flux of the form $G_4 = F_Y \wedge \omega_Y$. Here $\omega_Y$ is the 2-form dual to the zero-size $\mathbb P^1$ in the fiber over $S$ that corresponds to the node in the $SU(5)$ Dynkin diagram associated with the Cartan generator $T_Y$.

In presence of $U(1)_Y$ gauge flux the GUT matter decomposes into representations of $SU(3) \times SU(2) \times U(1)_Y$ as
\bea \label{splitting} 
 &   {\bf 24}        &\mapsto  ({\bf 8},{\bf 1})_{0_Y} + ({\bf 1},{\bf 3})_{0_Y} + ({\bf 1},{\bf 1})_{0_Y} + ({\bf 3},{\bf 2})_{5_Y}+  ({\bf \ov 3},{\bf 2})_{-5_Y},   \\ \nonumber
 &   \ov {\bf 5}\ \, &\mapsto  (\ov{\bf 3},{\bf 1})_{2_Y} + ({\bf 1},{\bf2 })_{-3_Y},   \\ \nonumber
 &   {\bf 10}        &\mapsto ({\bf 3},{\bf 2})_{1_Y} +  ({\bf \ov 3},{\bf 1})_{-4_Y} + ({\bf 1},{\bf 1})_{6_Y},   \\ 
 &   {\bf 5}_H       &\mapsto  ({\bf 3},{\bf 1})_{-2_Y} + ({\bf 1},{\bf2 })_{3_Y},\qquad  
    \ov {\bf 5}_H     \mapsto  (\ov {\bf 3},{\bf 1})_{2_Y} + ({\bf 1},{\bf2 })_{-3_Y}. \nonumber
\eea

The cohomology classes counting the MSSM matter contain factors of $L_Y^q$ with $q$ the $U(1)_Y$ charge of the state. In order to guarantee the same number of states within each $SU(5)$ family, the net $U(1)_Y$ flux through each matter curve must therefore vanish \cite{Beasley:2008kw,Donagi:2008kj}. As we will discuss in section \ref{proton}, absence of dimension 4 proton decay operators requires that the $\bf 5$ matter curve $P_{\bf 5}$ split into a ${\bf \ov 5_m}$ matter curve and a Higgs curve. Then what we need for absence of exotics in incomplete multiplets is
\bea
\label{cond-Y}
c_1(L_Y) \cdot P_{\bf 10} = 0 = c_1(L_Y) \cdot P_{\bf \ov 5_m}.
\eea
On the other hand, if in addition the Higgs curve splits into two curves $P_{H_u}$ and $P_{H_d}$, non-zero hypercharge flux through these Higgs curves allows for an elegant solution to the doublet-triplet splitting problem if it is chosen such as to  project out the massless  $( {\bf 3},{\bf 1})_{2_Y}$ within ${\bf \ov 5_H}$  \cite{Beasley:2008kw}.

Having outlined the general approach we next discuss three of the challenges which must be met for constructions with hypercharge flux breaking.
\subsubsection*{Massless $U(1)_Y$}

$U(1)_Y$ must remain massless after GUT symmetry breaking. In the language of IIB 7-branes with gauge flux, it is well-known that the Chern-Simons coupling (\ref{S-CS})  
leads to a St\"uckelberg mass for $U(1)_Y$ by dimensional reduction of the Ramond-Ramond four-form $C_4$ along 2-cycles on $S$,
 \bea
 \label{Stuck}
S_{\rm Stuckelberg} \simeq  \int_{\mathbb R^{1,3}}{F_Y^{4D} \wedge c_2^{(i)} \,  \, \,   {\rm tr} T_Y^2 \,  \int_S   c_1({L}_Y)  \wedge \i^* \omega_i  }.
\eea
 Here $ \omega_i  $ denotes a basis of $H^2(B, \mathbb Z)$ and we have decomposed $C_4 = c_2^{(i)} \wedge \omega_i$ with $c_2^{(i)}$ denoting two-forms in four dimensions. A mass term for $U(1)_Y$ can only be avoided if the gauge flux $F_Y$ is switched on exclusively along 2-cycles in $S$ which are homologically trivial as two-cycles in the ambient geometry \cite{Jockers:2004yj,Buican:2006sn};  in this case $c_1({L}_Y)$ is orthogonal to $\iota^{\ast} H^2(B,\mathbb Z) $.  Cycles of this type are said to lie in the relative cohomology of $S$ with respect to  $B$. Note that to ensure this topological constraint one needs full control of the global compactification geometry. This is because the question of triviality of a 2-cycle on $S$ can only be answered by studying an explicit embedding of $S$ into a compact geometry.

\subsubsection*{Absence of massless bulk exotics} 

The decomposition of the $SU(5)$ GUT matter displayed in (\ref{splitting}) contains the fields $({\bf 3},{\bf 2})_{5_Y}+  ({\bf \ov 3},{\bf 2})_{-5_Y}$. From the point of view of the MSSM these are exotic matter states which must be absent at the massless level for phenomenological viability of the model. 
Since they descend from the adjoint of $SU(5)$ they correspond to modes propagating along the entire GUT divisor $S$ and are therefore "bulk" states.  From the discussion  of bulk matter states in section \ref{matter-gen} we recall that these states are counted by the cohomology groups of the hypercharge flux $L_Y^q$ along $S$, where $q$ is the $U(1)_Y$ charge.
Absence of exotic states $({\bf 3},{\bf 2})_{5_Y}+  ({\bf \ov 3},{\bf 2})_{-5_Y}$ therefore requires that $H^i(S, { L_Y}^{\pm 5}) = 0$. 
Vanishing cohomology for such a high power of line bundles is difficult to engineer. A way out is to admit a suitably fractional line bundle ${\cal L}_Y$ instead of the integer quantised bundle $L_Y$ \cite{Beasley:2008kw}. A clean way to define this is by a twisting procedure that works both in perturbative Type IIB  \cite{Blumenhagen:2008zz} and in the $E_8$-based F-theory models \cite{Blumenhagen:2009up} discussed in these lectures. 
For such embeddings, the potential exotics are eventually counted by $H^i(S, {\cal L}_Y^{\pm 1})$, which therefore has to vanish. Here ${\cal L}_Y$ denotes the hypercharge bundle in the twisted embedding.
This constraint poses certain restrictions on the type of hypercharge flux switched on along the GUT brane. e.g.\ for a del Pezzo surface it can be achieved if and only if $c_1({\cal L}_Y) = E_i- E_j$ for $i \neq j$ \cite{Beasley:2008kw}.

As a final remark we note that it is this absence of bulk exotics that cannot be achieved for an analogous breaking of the GUT group $SO(10)$ by internal fluxes \cite{Beasley:2008kw}.
This is the technical reason why $SO(10)$ GUT models have received less attention in the F-theory literature. Alternatively, $SO(10)$ can be broken to flipped $SU(5)\times U(1)_X$ via suitable fluxes, and subsequently the standard field theoretic GUT Higgs mechanism \cite{Antoniadis:1987dx} can be invoked to break $SU(5) \times U(1)_X$ to the MSSM gauge group. For recent discussions of flipped $SU(5)$ in local and compact  F-theory setups see   e.g. \cite{Jiang:2009za,Li:2009fq} and, respectively,  \cite{Chen:2010ts,Chen:2010tp,Chung:2010bn}. An assessment of the phenomenological prospects of flipped $SU(5)$ and its realisations within F-theory has appeared in \cite{Kuflik:2010dg} and references therein.

\subsubsection*{Gauge coupling unification}

The most subtle and controversial challenge of the hypercharge GUT breaking scenario arises in the context of  gauge coupling unification \cite{Donagi:2008kj,Blumenhagen:2008aw}.
The problem is that - at least from a Type IIB perspective - hypercharge flux along  the GUT divisor seems to spoil the tree-level equality of the MSSM gauge couplings  at the GUT scale. To see this one must obtain the gauge kinetic function $f$ of the four-dimensional GUT gauge theory, defined as
\bea
S^{(4D)}_{YM} =   \frac{1}{2} \, {\rm Re}(f) \, \int_{{\mathbb R}^{1,3}}{\rm tr} \, F \wedge \star F    +  \frac{1}{2}\, {\rm Im}(f) \,  \int_{{\mathbb R}^{1,3}}  {\rm tr} \, F \wedge  F,
\eea
from the 7-brane effective action by dimensional reduction.
In Type IIB 7-brane language, the kinetic Yang-Mills and the topological Chern-Simons term follow from reduction of  $S_{DBI}$ and $S_{CS}$ in equ.\ (\ref{S-CS}), respectively.
The flux induced corrections to the leading order gauge kinetic function 
\bea
f_S  =  \frac{1}{g_s}  \frac{{ \rm Vol}_{S}}{\ell_s^4}  + i \int_{S} C_4
\eea
 can be deduced from the contribution to ${\rm Im}(f)$ encoded in the Cherns-Simons term proportional to $\int C_0 \wedge {\rm tr} F^4$.  To this end one takes into account holomorphicity of $f$ together with the fact that $C_0$ and $g_s$ combine into the holomorphic field $\tau =  C_0 + \frac{i}{g_s}$. What is important is that the contribution from hypercharge flux differs for the three gauge couplings $\alpha_s$, $\alpha_w$, $\alpha_Y$ of the MSSM gauge groups and distorts the gauge coupling relations at the Kaluza-Klein scale. The exact relation depends on the precise group theoretic embedding, see \cite{Blumenhagen:2008aw} and \cite{Donagi:2008kj} for two different types of embedding.
Final agreement on the interpretation of the physical consequences for unification has not yet been achieved in the literature. A conceptual difficulty is that a derivation of the critical $F^4$-term purely in F/M-theory and without reference to the weakly-coupled Type IIB language has not been provided so far.
In addition, threshold corrections from Kaluza-Klein and winding states have to be taken into account \cite{Donagi:2008kj} in a consistent manner. In particular ref. \cite{Conlon:2009qa} argues that the inclusion of the latter enhances the scale from where the gauge couplings run to the winding scale as a direct consequence of the fact that the hypercharge flux has to be trivial in the ambient space, as discussed around (\ref{Stuck}). 
The conclusions of \cite{Blumenhagen:2008aw} are that extra thresholds below the GUT scale are required to reconcile the flux-induced splitting of the gauge couplings with one-loop GUT unification. The minimal such threshold could be played by the unavoidable massive Higgs triplets  \cite{Blumenhagen:2008aw}, while more radical approaches consider incomplete multiplets of massive exotic matter \cite{Marsano:2009gv,Marsano:2009wr} (see also \cite{Leontaris:2009wi}).

\subsection{Proton decay}
\label{proton}

A classic topic in GUT model building is proton stability.
In particular, avoiding dimension 4 and dimension 5 proton decay operators is key to the phenomenological viability of $SU(5)$ models \cite{Murayama:2001ur,Raby:2002wc}.
We therefore face the question if string compactifications - here those of F-theory - add new ingredients to achieve these requirements. 
On the one hand, we can seek for stringy realisations of known field theoretic mechanisms such as favourable symmetries. In a more ambitious vein, new types of selection rules might become available in string theory that have no obvious four-dimensional counterpart. 

Let us begin with dimension 4 proton decay. 
Dangerous MSSM operators of the type $u_R^c \, d_R^c \, d_R^c$, $L \, L \, e_R^c$, $Q \, L \, d_R^c$ descend from a potential coupling  ${\bf 10 \, \ov 5_m \, \ov 5_m}$, which must therefore be prevented. The same holds for its unwanted cousin  ${\bf 10 \, \ov 5_H \, \ov 5_H}$. It was noted already in \cite{Beasley:2008kw} that a necessary condition for absence of ${\bf 10 \, \ov 5_m \, \ov 5_m}$ while allowing at the same time for the Yukawa couplings ${\bf 10 \, \ov 5_m \, \ov 5_H}$ and ${\bf 10 \, \bf 10 \,  5_H}$ is that the $\bf 5$ matter curve splits into (at least) two curves $P_m$ for $\bf \ov 5_m$ and $P_H$ for $\bf 5_H + \ov 5_H$.  From the perspective of the Weierstrass model, the splitting of matter curves corresponds to a non-generic situation that requires the restriction of some of the complext structure moduli, more on that in a second. If this splitting were sufficient to prevent dimension 4 proton decay, it would furnish an example of a geometric selection rule.

As found in \cite{Hayashi:2009ge}, however, typically $P_m$ and $P_H$ intersect at points without further singularity enhancements. If this happens, the wavefunction of $\bf \ov 5_m$ and $\bf 5_H + \ov 5_H$ obey the same boundary conditions and dangerous couplings of the type ${\bf 10 \, \ov 5_m \, \ov 5_m}$ are re-introduced.
To be on the safe side, explicit field theoretic selection rules have to be implemented. The minimal such selection rule - R-parity -  could descend from a geometric discrete ${\mathbb Z}_2$ symmetry that acts appropriately on the massless modes \cite{Hayashi:2009ge}, but no concrete realisations of this idea have been constructed as of this writing. More radically, a (massive) $U(1)$ selection symmetry can forbid unwanted Yukawa couplings \cite{Hayashi:2009ge}. The maybe simplest example of such a $U(1)$ symmetry is $U(1)_X$ with charge assignments \cite{Marsano:2009gv}
\beq
\label{charges}
{\bf 10}_1, \quad\quad  
({\bf \ov 5}_m)_{-3}, \quad\quad ({\bf 5}_H)_{-2} +  ({\bf \ov 5}_H)_{2},
\eeq
but other examples such as a Peccei-Quinn type $U(1)_{PQ}$ \cite{Marsano:2009wr} with different charge assignments for $H_u$ and $H_d$ or models with several $U(1)$s  \cite{Dudas:2010zb} have also been considered.

At the level of model building the implementation of such abelian symmetries requires a further specification of the complex structure moduli of the Weierstrass model. For global Tate models this  leads to so-called $U(1)$ restricted Tate models  \cite{GW-new}. In particular the engineering of, say, a $U(1)_X$ symmetry automatically leads to a split of the ${\bf \ov 5_m}$ and the Higgs curve. Note that as remarked before all questions associated with $U(1)$ symmetries defy a local treatment and require knowledge of the full compactification data of the 4-fold. This is intuitively clear because a  VEV of a $U(1)$ charged GUT singlet localised away from the GUT brane $S$ can higgs the abelian symmetry. Previously, the implementation of $U(1)$ symmetries had been studied via so-called split spectral covers \cite{Marsano:2009gv}.  Split spectral covers can be regarded as the restriction of $U(1)$ restricted Tate models to the neighbourhood of the GUT brane.
As such they are not sufficient to guarantee the presence of $U(1)$ symmetries \cite{Watari-new,GW-new}.

Dimension 5 proton decay can in principle be prevented via a missing partner mechanism \cite{Beasley:2008kw}  if also $\bf 5_H$ and $\bf \ov 5_H$ localise on separate curves. One possibility to achieve this is in the context of $U(1)_{PQ}$ extended models \cite{Marsano:2009wr}, where the abelian selection rule forbids dimension 5 proton decay. For the present-day realisations of this scenario, exotics in incomplete GUT multiplets result as a side-effect because of a tension of the condition (\ref{cond-Y}) with the precise structure of the matter curves. These exotics in turn affect gauge coupling unification as outlined in the previous section. Whether or not such incomplete multiplets are strictly unavoidable within the hypercharge GUT breaking framework if dimension 5 proton decay is to be prevented is still under investigation.

\subsection{Further developments}
\label{sec_Fur}

Our presentation of the phenomenological properties of F-theory GUTs has only covered some of the crudest aspects, and many more advanced phenomenological topics have been studied in the literature. For a review devoted specifically to the phenomenology of F-theory constructions and a more complete list of references we recommend \cite{Heckman:2010bq}.
Topics worth highlighting include these:
\begin{itemize}
\item
The local nature of brane models offers the possibility of studying the structure of Yukawa couplings without referring to the details of the global geometry.
Investigations of the Yukawa and flavour structure of $SU(5)$ GUT models in this context  and various phenomenological scenarios have appeared in \cite{Font:2008id,Heckman:2008qa,Hayashi:2009ge,Heckman:2009mn,Font:2009gq,Cecotti:2009zf,Conlon:2009qq,Hayashi:2009bt,Marchesano:2009rz,Dudas:2009hu,King:2010mq}. 
\item
Possible connections with  neutrino physics are the subject of \cite{Bouchard:2009bu,Tatar:2009jk}.
\item
On a more formal level, instanton effects in F-theory have been reconsidered in the recent literature \cite{Blumenhagen:2010ja,Cvetic:2009ah,Cvetic:2010rq,Donagi:2010pd} with special attention to the generation of phenomenologically viable matter couplings known from the weakly coupled Type II limit (see e.g.\ \cite{Blumenhagen:2009qh} for a review and references).
\end{itemize}

Another focus in the recent literature is the realisation of these and other model building ideas in concrete compact examples. The motivation behind this is, as was stressed already, that certain questions of phenomenological relevance cannot be disentangled from the global geometric structure. Compact F-theory GUT vacua have been constructed in \cite{Marsano:2009ym,Blumenhagen:2009yv,Marsano:2009gv,Krause,Marsano:2009wr,Chen:2010ts}. 
Methods of toric geometry allow one to explicitly construct fully-fledged singular Calabi-Yau 4-folds and their explicit resolution as in \cite{Blumenhagen:2009yv,Krause,Chen:2010ts}  in a way that keeps full control of the singularities of the Tate model.
 As far as the construction of gauge flux is concerned, the models \cite{Marsano:2009ym,Blumenhagen:2009yv,Marsano:2009gv,Krause,Marsano:2009wr,Chen:2010ts} rely on the spectral cover approach outlined in section \ref{spectral_cover}. Modulo the caveats pointed out there, 3-generation GUT models have been achieved.

 \vspace{10pt}

\vspace{30pt}

 {\noindent  {\Large \bf Acknowledgements}}
 \vskip 0.5cm 

I would like to thank the organisers of the CERN Winter School on Supergravity, Strings and Gauge Theory 2010 for the invitation to lecture, and the participants  for their interest and their valuable questions and remarks. 
I am indebted to my collaborators over the past couple of years who have been sharing my interest in F-theory model building: R. Blumenhagen, B. Jurke, T. Grimm and S. Krause. I thank M. Kerstan and  S. Krause for useful comments on a draft of these lecture notes.
Finally I am grateful to the Max-Planck-Institute in Munich, the KITP Santa Barbara and Ecole Normale, Paris, for hospitality during parts of this work. This project was supported in part by the SFB-Transregio 33 ``The Dark Universe'' by the DFG  and the National Science Foundation under Grant No. PHY05-51164.

\clearpage
\nocite{*}
\bibliography{rev-CERN}
\bibliographystyle{utphys}

\end{document}